\documentclass[iop]{emulateapj}

\usepackage{pslatex}
\usepackage{amsmath}
\usepackage{textcomp}

\usepackage[dvips]{hyperref}
\hypersetup {
    bookmarks=true,                    
    pdftitle={Self-Consistent Magnetic Stellar Evolution Models of the Detached, Solar-Type Eclipsing Binary EF Aquarii},                   
    pdfauthor={Gregory A. Feiden},     
    pdfsubject={},                     
    colorlinks=true,                   
    linkcolor=blue,                   
    citecolor=blue,                    
    urlcolor=blue                      
}
\newcommand{\myemail}{Gregory.A.Feiden.GR@Dartmouth.edu}
\newcommand{\bcemail}{Brian.Chaboyer@Dartmouth.edu}

\newcommand{\thermt}[4] {\left(\frac{d #1}{d #2}\right)_{#3,#4}}
\newcommand{\thermd}[4] {\left(\frac{\partial #1}{\partial #2}\right)_{#3,#4}}
\newcommand{\tgrad}{\nabla_{\textrm{temp}}}

\shorttitle{Magnetic Stellar Models of EF Aquarii}
\shortauthors{Feiden \& Chaboyer}
\slugcomment{Submitted Draft: Version 2; \today}

\begin{document}


\title{Self-Consistent Magnetic Stellar Evolution Models of the Detached,\\ Solar-Type
Eclipsing Binary EF Aquarii}

\author{Gregory A. Feiden\altaffilmark{1} and Brian Chaboyer}
\affil{Department of Physics and Astronomy, Dartmouth College, 6127 Wilder Laboratory, Hanover, NH 03755, USA; \href{mailto:\myemail}{\myemail}, \href{mailto:\bcemail}{\bcemail}}
\altaffiltext{1}{Neukom Graduate Fellow.}


\begin{abstract}
We introduce a new one-dimensional stellar evolution code, based on the existing Dartmouth
code, that self-consistently accounts for the presence of a globally pervasive magnetic
field. The methods involved in perturbing the equations of stellar structure,
the equation of state, and the mixing-length theory of convection are presented
and discussed. As a first test of the code's viability, stellar evolution models 
are computed for the components of a solar-type,
detached eclipsing binary (DEB) system, EF~Aquarii, shown to exhibit large disagreements
with stellar models. The addition of the magnetic perturbation corrects the radius 
and effective temperature discrepancies observed in EF~Aquarii. Furthermore, the 
required magnetic field strength at the model photosphere is within a factor 
of two of the magnetic field strengths estimated from the stellar
X-ray luminosities measured by \emph{ROSAT} and those predicted from Ca {\sc ii}
K line core emission. These models provide firm evidence 
that the suppression of thermal convection arising from the presence of a
magnetic field is sufficient to significantly alter the structure of 
solar-type stars, producing noticeably inflated radii and cooler effective
temperatures. The inclusion of magnetic effects within a stellar 
evolution model has a wide range of applications, from DEBs and exoplanet 
host stars to the donor stars of cataclysmic variables. 
\end{abstract}


\section{Introduction}
\label{sec:intro}

The vast array of physics incorporated in standard low-mass stellar evolution models
\citep[see e.g.,][and references therein]{CB1997,BCAH98,Dotter2007,Dotter2008} appears to 
be insufficient for predicting the properties of low-mass stars. Studies of 
detached eclipsing binary (DEB) systems allow for a very precise determination
of the mass and radius of the individual stellar components with uncertainties
commonly below 2\%. Over the past 
two decades, these precision studies have accumulated strong evidence that the radii
predicted by low-mass stellar evolution models are deflated compared to
the observations, at a given mass \citep[e.g.,][]{Popper1997,Torres2002,Ribas2006,
Morales2009,Torres2010,Kraus2011}. Typically quoted is that DEB radii appear
to be systematically 10\% larger than model predictions and DEB effective
temperatures are 5\% cooler than the models imply. Further evidence has 
been garnered by studies of single low-mass stars \citep{Berger2006, Morales2008},
which confirm the aforementioned trends. Although, \citet{Boyajian2012}
present results that may be interpreted as counter to these claims.

Recent work has shown that the disagreements may not be as severe 
as previously believed, when the age and metallicity of the DEBs are considered
\citep{FC12}. Still, many DEB systems display moderately inflated radii (less than 5\%)
with a small subset displaying radically inflated radii (upward of
10\%) compared to stellar models \citep{FC12,Terrien2012}. This modest radius offset 
between observations and models and the presence of strongly inflated stars,
suggests that stellar evolution models must invoke new physics to account 
for the appearance of inflated radii.

Implicated as the culprit for the observed inflated radii and cooler effective
temperatures is magnetic activity \citep{Ribas2006, lopezm2007,Morales2008}. 
The systems at the heart of the problem are often close binaries with short 
orbital periods. Tidal synchronization of the components acts to spin-up 
the rotation rate of each star, enhancing the dynamo mechanism and thus 
supporting a stronger magnetic field within each. While there are a number
of direct magnetic field measurements for single stars, there are few 
among fast rotating binaries. Instead, the hypothesis is
reinforced by the presence of strong chromospheric H$\alpha$ emission 
\citep{Morales2008, Stassun2012}, chromospheric Ca~{\sc ii}~H and K 
emission \citep{Skumanich1975}, as well as coronal X-ray emission \citep{lopezm2007} among
inflated stars. The presence of such emission features is considered to be 
the result of the dissipation of magnetic energy in the stellar atmosphere.

There are, however, stars from DEBs that display inflated radii, despite
existing in long-period systems. Both \object{LSPM J1112+7626} \citep{Irwin2011}
and \object{Kepler-16} \citep{Doyle2011,Winn2011,Bender2012} have orbital periods of about 41
days, suggesting that the components are tidally unaffected by the presence of
their companion. It is possible that these stars have not had sufficient time
to shed angular momentum \citep{Skumanich1972}, preserving strong magnetic fields that they likely
possessed near the zero-age main sequence. Whether the low-mass stars in these systems are still
magnetically active enough to affect the stellar radius remains unclear.
The work by \citet{Winn2011} appears to suggest that Kepler-16 is still relatively
young (2 -- 4 Gyr) as the primary is more active than the Sun, based on
Ca {\sc ii} line emission. This supports the notion that magnetic fields may
be influencing the structure of each component. On the other hand, the age of LSPM J1112+7626, 
estimated from gyrochronology, reveals that it is approximately 9~Gyr old 
\citep{Barnes2010}, supporting the notion that magnetic activity is likely 
not playing a prominent role in either star's evolution.

Further complicating our picture of low-mass DEBs, \object{KOI-126} \citep{Carter2011} 
curiously matches standard stellar evolution models \citep{Feiden2011,SD12}.  
The mass of KOI-126 B and C and their orbital period are very similar to 
\object{CM Draconis}, one of the quintessential low-mass DEB 
systems displaying inflated radii \citep{Lacy1977,Morales2009,Terrien2012}. 
It is interesting,
then, that stellar evolution models would even come close to accurately 
predicting the observed stellar properties of KOI-126. 
Given what is known about CM~Draconis, one would expect KOI-126 to
display inflated radii as a consequence of moderate magnetic activity. 
Although, \citet{MM12} speculate that KOI-126 should, in fact, not be 
terribly active and find it unsurprising that standard models should 
fit the system so well. 

Despite the identification of a potential culprit responsible for inflating
the radii of DEB components, only ad hoc procedures for treating 
the effects of magnetic fields have been introduced \citep{MM01,Chabrier2007}. 
The method examined by \citet{Chabrier2007} included 
artificially decreasing the convective mixing-length parameter, so as to 
mimic the effect of a global magnetic field within the star, as well 
as artificially reducing the star's bolometric flux in an effort to reproduce
the effects of photospheric spots. A second method, proposed by \citet{MM01}, 
altered the Schwarzschild criterion by perturbing the adiabatic gradient 
in a manner consistent with the work of \citet{GT66}. 

Investigations by both groups appear to be at odds with 
one another. \citet{Chabrier2007} and \citet{Morales2010} claim that starspots
appear to be the dominate mechanism inflating stellar radii, and that modifications 
to convection require unrealistic magnetic field strengths (i.e., reductions
in the mixing length in their formulation). On the other hand, \citet{MM01} 
and \citet{MM12} conclude the opposite
that reduction in convective efficiency is ultimately the dominant mechanism.
Regardless of which is really the dominant mechanism, both approaches are 
inherently ad hoc, yet both are capable of reproducing the observed 
inflated stellar radii.

In this paper, we introduce a self-consistent treatment of a globally pervasive magnetic 
field embedded in the framework of the Dartmouth stellar evolution code 
\citep{Dotter2007,Dotter2008}. Our approach follows the outline provided 
by \citet{ls95}, though we deviate from their method in a number of ways
that are described below. All of the stellar structure equations, including 
those in the equation of state, are self-consistently modified, as opposed
to arbitrarily altering a single quantity. In this way, modifications to the efficiency
of thermal convection are accounted for in a more complete fashion, owing
to the full thermodynamic treatment of the magnetic field. Overall, the 
approach used to model magnetic effects can be considered analogous to 
the parameterized mixing-length treatment of convection.\footnote{In so far
as reducing an inherently nonlinear, three-dimensional process into terms suitable
for a one-dimensional model.}
The viability of the models is tested against results from a recent study
 that characterized the DEB EF~Aquarii \citep{vos12}.

EF~Aquarii (\object{HD 217512}; henceforth EF~Aqr) is a solar-type DEB found to contain two 
components displaying drastically inflated radii \citep{vos12}. Fundamental
parameters of the system are quoted in Table~\ref{tab:efaqr}. What is most striking, 
is the similarity of the secondary to the Sun and the entire system to 
\object[*alf Cen]{$\alpha$ Centauri} A and B, in terms of the stellar masses and composition.
Although the secondary appears similar to both $\alpha$ Cen B and the Sun,
its radius appears to be about 10\% larger than one would expect based
on stellar evolutionary calculations. The effective
temperature of the primary further reveals that both components suffer 
from substantial radius inflation.

\renewcommand{\arraystretch}{1.1}
\begin{deluxetable}{l c c}
    \tablewidth{0.8\columnwidth}
    \tablecolumns{3}
    \tablecaption{Fundamental Stellar Parameters for EF Aqr}
    \tablehead{
        \colhead{Parameter} & \colhead{EF Aqr A} & \colhead{EF Aqr B}   
        }
    \startdata
    $M\, (M_{\odot})$ & 1.244$\pm$0.008 & 0.946$\pm$0.006  \\
    $R\, (R_{\odot})$ & 1.338$\pm$0.012 & 0.956$\pm$0.012  \\
    $T_{\textrm{eff}}$ (K) & 6150$\pm$65  & 5185$\pm$110 \\
    $[{\rm Fe/H}]$  & \multicolumn{2}{c}{0.00$\pm$0.10}
    \enddata
    \label{tab:efaqr}
\end{deluxetable}

In an effort to reconcile the observations with predictions from theoretical
models, \citet{vos12} reduced the value of the convective mixing-length. 
They found 
$\alpha_{\textrm{MLT}}$ of 1.30 and 1.05 were required for the primary
and secondary, respectively, compared to their solar-calibrated value of 
1.68. They concluded that fine-tuning the models allows
for an accurate description of the observed properties.

Reduction of the required convective mixing length may be physically motivated 
in two ways: (1) naturally inefficient convection and (2) magnetically
suppressed convection. While we must be careful to not read too much into 
the reality of mixing-length theory, in stellar evolution models the 
mixing length is an intrinsic ``property'' of convection. Thus, reducing
the mixing length is akin to saying convection is not very efficient at
transporting excess energy.

\citet{Bonaca2012} calibrated the convective mixing length for solar-like
stars using asteroseismic results provided by the \emph{Kepler Space Telescope}.
They found that the value of $\alpha_{\rm MLT}$ in stellar models is
tied to stellar properties (i.e., $\log g$, $\log T_{\rm eff}$, and
[$M$/H]). Applying the \citet{Bonaca2012} empirical calibration to the stars
in EF~Aqr, we find that the primary and secondary component require 
$\alpha_{\rm MLT}$~=~1.68 and 1.44, respectively. Again, compared with 
their solar calibrated value of $\alpha_{\rm MLT}$~=~1.68.
The asteroseismically adjusted mixing-lengths are significantly larger than the
fine-tuned values determined by \citet{vos12}. Therefore, it appears that 
naturally inefficient convection is insufficient to explain the inflated 
radii of EF~Aqr. We are left with the option that magnetic fields may possibly
be to blame.

In what is to follow, we describe a self-consistent approach
to modeling the effects of a globally pervasive magnetic field with application
to the EF~Aqr system. Details of the standard Dartmouth models are presented 
in Section~\ref{sec:std_model} followed by a description of the magnetic
perturbations introduced to the code in Sections~\ref{sec:mag} and 
\ref{sec:num_test}. Section~\ref{sec:results} demonstrates the ability 
of the invoked perturbations to reconcile the models with the observations.
We conclude with a further discussion of the results and their implications
in Section~\ref{sec:disc}.


\section{Models}
\label{sec:std_model}
The framework throughout which magnetic effects are invoked is provided by the 
Dartmouth Stellar Evolution Program \citep[DSEP;][]{Dotter2008},\footnote{Available at: 
\url{http://stellar.dartmouth.edu/models/}}
a descendant of the Yale Rotating Evolution Code \citep{Guenther1992}. 
Standard stellar evolution henceforth refers to the basic physics without 
any magnetic perturbation. The standard physics incorporated
in DSEP has been described previously \citep{Chaboyer1995,
Chaboyer2001,Bjork2006, Dotter2007,Dotter2008,Feiden2011}, although we will
briefly review the elements that are pertinent for the current study.

Above $0.80\, M_{\sun}$, DSEP invokes an ideal gas equation of state (EOS)
supplemented by a Debye-H\"{u}ckel correction in order to account for 
ion-charge shielding \citep{Chaboyer1995}. This EOS is computed analytically,
in a self-consistent manner, within the code and does not rely on the 
interpolation within EOS tables. Opacities are drawn from two sources, the
OPAL opacities for the high temperature regime \citep{Iglesias1996} 
complimented by the Ferguson low-temperature opacities \citep{Ferguson2005}.
Surface boundary conditions are defined using the {\sc phoenix ames-cond} model
atmospheres \citep{Hauschildt1999a,Hauschildt1999b}, attached to the model
interior where $T = T_{\rm eff}$. 

Atomic diffusion and the gravitational settling of helium and heavy elements
are implemented using the prescription of \citet{Thoul1994}. Additional 
diffusion effects associated with turbulent mixing \citep{Richard2005} are
also included. Details of the latter are presented in \citet{Feiden2011}.
Finally, convective core overshoot is treated following the methods outlined
by \citet{Demarque2004}.

Required before any analysis pertaining to stellar evolution models is 
calibrating the model properties to the Sun. Determination of the initial 
solar helium and heavy element mass fractions along with a compatible 
mixing-length parameter ($Y_{\rm init}$, $Z_{\rm init}$, and $\alpha_{\rm MLT}$, 
respectively) was performed by calibrating a 1 $M_{\odot}$ model to the 
Sun. Our solar model was required to reproduce the solar radius, solar 
luminosity, radius at the base of the convection zone, and the solar 
photospheric ($Z/X$) at the solar age \citep[4.57 Gyr;][]{Bahcall2005}. 
The final set of parameters necessary to satisfy the above criteria for 
the \citet{GS98} solar composition was $Y_{\rm init} = 0.27491$, $Z_{\rm init} = 0.01884$, 
and $\alpha_{\rm MLT} = 1.938$.


\section{Magnetic Perturbation}
\label{sec:mag}

\subsection{Magnetic Field Characterization}
\label{sec:mag_character}
Investigating the effects of a global magnetic field on the interior
structure of a star over long time baselines, requires formulating a purely
three-dimensional (3D) phenomenon in terms suitable for a one-dimensional 
(1D) numerical model. Unfortunately, full 3D magnetohydrodynamic (MHD) models
are not yet capable of modeling stellar magnetic fields over the long time baselines 
required for stellar evolutionary calculations. This is in part due
to the rapid\footnote{Relative to a typical stellar lifetime.} diffusion of 
the magnetic field and the immense computational 
time required. Therefore, in order to probe the effects of a magnetic field,
we seek to avoid directly solving the induction equation
\begin{equation}
\frac{\partial {\bf B}}{\partial t} = \nabla \times \left({\bf u} 
 \times {\bf B}\right) + \eta \nabla^{2}{\bf B}.
 \label{eq:ind}
\end{equation}

While not actively seeking a solution to the full suite of 3D MHD equations, 
it is possible to use the theoretical framework of MHD to provide a reasonably
accurate 1D description of a magnetic field and its associated properties.
Ultimately, we are able to describe a magnetic field in terms of the MHD equations 
and then project out the radial component, the component necessary for stellar
evolutionary model computations.

The spatial and temporal evolution of a given magnetic field are governed,
quite naturally, by Maxwell's equations,
\begin{eqnarray}
\nabla \cdot {\bf E} & = & 0 \\
\nabla \times {\bf E} & = & -\frac{1}{c}\frac{\partial {\bf B}}{\partial t} \\
\nabla \cdot {\bf B} & = & 0 \\
\nabla \times {\bf B} & = & \frac{4\pi}{c}{\bf J} \label{eq:amp}
\end{eqnarray}
where within the stellar plasma, we assume any regions of excess charge inducing an electric potential 
will rapidly neutralize owing to the mobility of other charges (Debye shielding). 
Thus, we can safely assume that the plasma is electrically neutral, $\rho_e = 0$. For simplicity,
we here made another assumption, that temporal variations of the large-scale field
are small, suggesting that the conduction current dominates the displacement
current.

Now, let us consider the interactions between the electric and magnetic fields 
within a dense, ionized fluid moving with arbitrary velocity, {\bf u}. For
slow temporal evolution, non-relativistic dynamics may be described
by a single conducting fluid that obeys the classical equations of hydrodynamics 
coupled with the equations of electromagnetism; the MHD equations \citep{Jackson1999}. 
Considering a perfectly
conducting, non-viscous, non-rotating, compressible fluid in the presence of a
gravitational field, the MHD equations governing the system are Ohm's law for
a moving fluid,
\begin{equation}
{\bf J} = \sigma \left( {\bf E} + \frac{{\bf u} \times {\bf B}}{c}\right) ,
\end{equation}
the equation of mass continuity,
\begin{equation}
\frac{\partial \rho_m}{\partial t} + \nabla \cdot (\rho_m {\bf u}) = 0 ,
\end{equation}
where $\rho_m$ is the mass density, and the fluid equation of motion,
\begin{equation}
\rho_m\frac{d{\bf u}}{dt} = \frac{{\bf J} \times {\bf B}}{c} - \nabla
  \cdot \overleftrightarrow{{\bf P}} + \rho_m {\bf g}
\label{eq:eom1}
\end{equation}
with {\bf g} being the gravitational field vector and $\overleftrightarrow{{\bf P}}$ 
representing the gas pressure tensor. The electromagnetic term in the fluid
equation of motion is associated with the assumption that a magnetic field 
permeates the plasma. However, we have neglected forces associated with any
electric fields, for reasons detailed above.

Since, a priori, we have no knowledge of the current density within
a given fluid, we replace the current density within Equation~(\ref{eq:eom1})
using Equation~(\ref{eq:amp}). The equation of motion may now be written as
\begin{equation}
\rho_m\frac{d{\bf u}}{dt} = \frac{1}{4\pi} \left(\nabla \times {\bf B}\right) 
  \times {\bf B} - \nabla \cdot \overleftrightarrow{{\bf P}} + \rho_m {\bf g}.
\end{equation}
With the aid of a vector operation identity and knowing that the magnetic field
is divergenceless, this may again be rewritten as
\begin{equation}
\rho_m\frac{d{\bf u}}{dt} = \frac{1}{4\pi}({\bf B}\cdot\nabla){\bf B} - 
 \frac{1}{8\pi}\nabla B^2  - \nabla \cdot \overleftrightarrow{{\bf P}} 
 + \rho_m {\bf g}.
\end{equation}
Immediately, we recognize that the electromagnetic contributions on the right-hand
side are the familiar magnetic tension and pressure terms, respectively.
However, the final form of the equation of motion requires one further step. 
The first two terms on the right-hand side may be expressed as the divergence
of a magnetic stress tensor \citep{GB2005}, let us call it $\overleftrightarrow{{\bf T}}$, such that
\begin{equation}
\overleftrightarrow{{\bf T}} = -\frac{{\bf B}{\bf B}}{4\pi} + 
         \overleftrightarrow{{\bf I}} \frac{B^2}{8\pi}
\end{equation}
with $\overleftrightarrow{{\bf I}}$ representing the identity tensor. This
definition then implies,
\begin{equation}
\rho_m\frac{d{\bf u}}{dt} = -\nabla\cdot\left(\overleftrightarrow{{\bf T}} + 
 \overleftrightarrow{{\bf P}}\right) + \rho_m {\bf g}.
\label{eq:eom}
\end{equation} 
The magnetic stress tensor introduced above can be thought of an anisotropic 
pressure tensor, where the pressures it describes are intrinsic properties of 
the magnetic field.

Stars, however, are considered to be in hydrostatic equilibrium. This implies
the absence of bulk fluid motion, forcing the left-hand side of the equation
of motion to vanish. Therefore,
\begin{equation}
\nabla\cdot\left(\overleftrightarrow{{\bf T}} + 
 \overleftrightarrow{{\bf P}}\right) = \rho_m {\bf g},
 \label{eq:hse}
\end{equation}
which is a statement of magnetohydrostatic equilibrium.


\subsection{Stellar Structure Perturbations}
\label{sec:structure}
At the most fundamental level, one-dimensional stellar evolution codes simultaneously solve
a set of four coupled, first-order differential equations.\footnote{There 
are additional equations often included
to account for atomic diffusion. While included in DSEP, we do not seek
perturbations to these equations at the present time. See \citet{MZ05} for 
a rigorous treatment of mixing associated with magnetic fields.} They are the 
equation of mass conservation, hydrostatic equilibrium, energy transport,
and energy conservation. Qualitatively, we can easily predict how these 
equations will be altered by the presence of a magnetic field which may then
be translated into a quantitative description.

The equation of mass conservation should be unaltered by any magnetic perturbation.
Of course, this is assuming that mass removed by stellar winds is negligible
and that transient events that may remove mass (i.e., flares, coronal mass
ejections) are neglected. The stated conditions hold for our approach. Thus, 
\begin{equation}
\frac{d r}{d m} = \frac{1}{4\pi r^2 \rho}
\end{equation}
where we have dropped the subscript $m$ on the density and assume all references
to density are specifically to the mass density, unless otherwise noted.

Hydrostatic equilibrium, as we saw earlier in Equation~(\ref{eq:hse}), is 
modified through the inclusion of the magnetic pressure and tension. Projecting
out the radial component of the magnetic pressures, we are able to adapt 
the three-dimensional concept for one-dimensional models. Therefore, we have
\begin{equation}
\frac{d P}{d m} = -\frac{G m}{4\pi r^4} + \frac{1}{4\pi r^2 \rho}
 \left[\frac{({\bf B \cdot \nabla}){\bf B}}{4\pi} - {\bf \nabla} \left(
 \frac{B^2}{8\pi}\right)\right]\cdot {\bf \hat{r}}.
\end{equation}
The precise handling of the vector magnetic field within the code will 
be discussed later.

The final form of the energy transport 
equation is the same as if there were no perturbation. Namely,
\begin{equation}
\frac{d T}{d m} = \frac{T}{P}\tgrad \frac{dP}{dm}
\label{eq:et}
\end{equation}
where $\nabla_{\textrm{temp}}$ is the local temperature gradient. Magnetic 
perturbations to the stellar structure equations will self-consistently
alter the temperature gradient through various thermodynamic 
considerations. Of greatest importance will be the affects on the treatment
of convection. The full treatment will be discussed in the next subsection. 

Finally, there are changes to the parameters present in the canonical equation
of energy conservation in stellar evolution. Modifications to these parameters 
arise from the treatment of the specific thermodynamic equations (discussed in 
the next subsection) and additional terms that are electromagnetic in origin.
The final form of the energy conservation equation is
\begin{equation} 
\frac{d L}{d m} = \epsilon - \frac{d U}{d t}
  + \frac{P}{\rho^2}\frac{d \rho}{d t} + \frac{Q_{\rm ohm}}{\rho}
  + \frac{F_{\rm Poynt}}{\rho}.
\end{equation}

Aside from the first three standard terms on the right-hand side, there are
two additional electromagnetic terms. First, there is a Poynting flux 
associated with the field,
\begin{equation}
F_{\textrm{Poynt}} = \frac{c}{4\pi}{\bf E \times B},
\end{equation}
although as discussed above, we assume the $E$-field is zero everywhere. Next,
energy is also associated with the Ohmic dissipation of electric currents 
brought about by the resistive nature of the plasma. 
\begin{equation}
Q_{\textrm{ohm}} \propto I^2 R,
\end{equation}
where $I$ is the electric current and $R$ is the resistance of the medium.
Here, electrical currents are converted to heat that then is transmitted to
the surrounding plasma. Since we have assumed an infinitely conducting 
plasma, this energy term goes immediately to zero.


\subsection{Thermodynamic Considerations}
\label{sec:thermo}
The effects of a global magnetic field are introduced into the thermodynamic
framework supplied by DSEP following the approach outlined by \citet[hereafter LS95]{ls95}. 
Providing a detailed, step-by-step guide of the magnetic perturbation to 
the various thermodynamic quantities would prove tedious. Therefore, we
refer the reader to LS95 for a full derivation of each equation presented
below. Our aim in this subsection is to adequately summarize the pertinent aspects of 
LS95 and highlight where we diverge from their original approach.

At the core of the LS95 method is the specification of a new thermodynamic
state variable, $\chi$, such that 
\begin{equation}
\chi = \chi(r,\, \rho) = \frac{U_{\chi}}{\rho} = \frac{B(r)^2}{8\pi\rho}.
\label{eq:chi}
\end{equation}
The state variable $\chi$ is the magnetic energy per unit mass and $B(r)$
is the magnetic field strength at radius $r$. Unlike LS95,
our definition of $\chi$ depends on the radial distribution of the magnetic 
field strength and also on the density of the stellar plasma. Originally,
LS95 favored a mass-depth-dependent function, $\chi(M_{r})$. However, we 
moved away from this prescription when we realized several thermodynamic 
derivatives became divergent. Once the 
magnetic field strength is specified throughout the star, it is straightforward 
to calculate $\chi$ at each point within the model. 

The energy associated with the magnetic field arises due to forces exerted
by the magnetic field on the plasma. These forces are represented by the 
anisotropic pressure tensor present in Equation~(\ref{eq:eom}). As a first 
approximation, we convert the pressure tensor to a scalar pressure by taking
the trace of the pressure tensor to yield the mean magnetic pressure,
\begin{equation}
\langle P_{\textrm{mag}}\rangle \sim \frac{1}{3}\textrm{Tr}
  \left( -\frac{{\bf B}{\bf B}}{4\pi} + \overleftrightarrow{{\bf I}} \frac{B^2}{8\pi} \right).
\end{equation}
Since we are not solving the full set of MHD equations, we look, instead,
to set approximate upper and lower limits on the scalar pressure.
Assuming a Cartesian coordinate system, if we imagine the magnetic field 
is parallel to the $z$-axis, or for a star, the rotational axis, then we 
may expand the pressure tensor to read
\begin{equation}
    \overleftrightarrow{{\bf T}} = \left[
    \begin{array}{c c c}
        B^2/8\pi & 0 & 0 \\
        0 & B^2/8\pi & 0 \\
        0 & 0 & -B^2/4\pi + B^2/8\pi
    \end{array}
    \right].
\end{equation}
Note that there is an isotropic magnetic pressure associated with each diagonal 
element along with the additional magnetic tension term in the final element.
Since tension is directed along the field line, the tension exists in
the $z$-direction only, in this instance. Taking the trace, we find
\begin{equation}
\langle P_{\textrm{mag}}\rangle = \frac{1}{3}\left(\frac{B^2}{8\pi}\right) = \frac{1}{3}\chi\rho.
\end{equation}

The above equation is satisfied for a magnetic field where a strong tension 
component is present. However, if we assume that there is no tension at all,
then, following the same procedure, 
\begin{equation}
\langle P_{\textrm{mag}}\rangle = \frac{1}{3}\left(\frac{3B^2}{8\pi}\right) = \chi\rho.
\end{equation}
We can now limit the strength of the scalar magnetic pressure within the 
one-dimensional framework. Specifically,
\begin{equation}
\frac{1}{3}\chi\rho \le \langle P_{\textrm{mag}}\rangle \le \chi\rho.
\end{equation}
Defining a ``geometry parameter,'' akin to LS95, allows us to emulate the effects of having
a strongly curved field or a field with no curvature, and varying degrees 
between the two extremes. This geometry parameter is defined such that we 
recover the average magnetic pressure for each case above,
\begin{equation}
\langle P_{\textrm{mag}} \rangle = \left(\gamma - 1\right)\chi\rho 
  \equiv P_{\chi}
\end{equation}
where 
\begin{equation}
\gamma = \left\{ 
  \begin{array}{c c l}
    2   & & \textrm{tension-free} \\
    4/3 & & \textrm{maximum tension}
  \end{array}
  \right. .
\end{equation}
In both cases, the appropriate expression for the scalar magnetic pressure
is returned. With the magnetic energy density and pressure formulated as 
scalars, we have successfully converted the inherently three-dimensional magnetic field 
into a one-dimensional magnetic perturbation. In the process, we have also reproduced the 
scalar parameters originally presented by LS95.


\subsubsection{Equation of State}
\label{sec:eos}
The derivations that follow hereafter in Sections~\ref{sec:eos} and \ref{sec:mlt} 
are provided as a review of the LS95 method to enable transparency and enhance
the clarity of discussions concerning the application of our models. 
Original, complete derivations are to be found in LS95. We do, however, deviate from 
their paper in Equation~(\ref{eq:longest}), where it is stated explicitly below.

We have just seen that magnetic fields exert forces on the plasma and, thus,
carry an associated pressure, tension, and energy. The introduction of these
terms into the equations of stellar structure then necessitates the inclusion
of these parameters in the EOS of the system. Again, 
we will mention only the most important modifications, deferring to LS95 
for a rigorous treatment. Beginning with the first law of thermodynamics, 
\begin{equation}
dQ = TdS = dU + PdV
\end{equation}
we recognize that each term contains, now, both the standard gas and radiation
terms as well as a new magnetic contribution,
\begin{equation}
dQ = T(dS_0 + dS_{\chi}) = (dU_0 + dU_{\chi}) + P_0dV.
\end{equation}
In the above equation, the magnetic perturbation rightly does not contribute
any work. However, in order to write the equation as a function of the total
pressure, we may subtract off the magnetic contribution,
\begin{equation}
TdS = dU + PdV - (\gamma - 1)\frac{\chi}{V}dV
\label{eq:firstlaw}
\end{equation}
where we take the volume to be the specific volume, $V = \rho^{-1}$. Hereafter,
it is also assumed that any unsubscripted quantity refers to the total quantity
while gas and radiation are lumped under the subscript 0 (zero) convention and magnetic
variables carry a subscript $\chi$.

Equation~(\ref{eq:firstlaw}) is the new ``non-standard'' first law of thermodynamics
and suggests that 
\begin{equation*}
V \textrm{ or } \rho = f(P, \, T, \, \chi)
\end{equation*}
and 
\begin{equation*}
U \textrm{ or } S = f(\rho, \, T, \, \chi).
\end{equation*}
Following the derivation by LS95, explicitly writing out the other state
variables illustrates the effects of adding the magnetic perturbation. 
Ignoring constants for clarity and ease,
\begin{eqnarray}
P & = & \rho T + \frac{1}{3}T^4 + \chi\rho (\gamma - 1) \\
\rho & = & \left[P - T^4/3\right]/\left[T + (\gamma - 1)\chi\right] \\
U & = & \frac{3}{2}T + \frac{T^4}{\rho} + \chi
\end{eqnarray}
which are all subject to the EOS,
\begin{equation}
\frac{d\rho}{\rho} = \alpha \frac{dP}{P} - \delta \frac{dT}{T} - 
  \nu \frac{d\chi}{\chi}.
\end{equation}
The coefficients in the EOS above are defined as follows,
\begin{eqnarray}
\alpha & = & \thermd{\ln \rho}{\ln P}{T}{\chi} \\
\delta & = & -\thermd{\ln \rho}{\ln T}{P}{\chi} \\
\nu & = & -\thermd{\ln \rho}{\ln \chi}{P}{T}
\end{eqnarray}
and will be referred to, as such, throughout the rest of the paper. Note, 
that $\alpha$ carries no subscript and should not be confused with the 
convective mixing-length parameter, $\alpha_{\textrm{MLT}}$.

An immediate consequence of altering the thermodynamic variables is the 
effect on the specific heats,
\begin{eqnarray}
c_P & = & \thermt{Q}{T}{P}{\chi} = \thermt{U}{T}{P}{\chi} + \left[P - (\gamma-1)\chi\rho\right]\thermt{V}{T}{P}{\chi} \\
c_V & = & \thermt{Q}{T}{V}{\chi} = \thermt{U}{T}{V}{\chi}
\end{eqnarray}
which are related to one another via the relation,
\begin{equation}
c_P - c_V = \frac{P\delta^2}{\rho T \alpha}.
\end{equation}
The difference in specific heats is written just as it would be without
any magnetic perturbation, however, each term is self-consistently modified
by the presence of a magnetic perturbation.

The change in heat as a result of the magnetic perturbation follows and
is found to be
\begin{equation}
dQ = c_PdT - \frac{\delta}{\rho}dP + \left(\frac{P\delta\nu}{\rho\alpha\chi}
   + 1\right)d\chi,
\end{equation}
although, the addition of the purely magnetic term, $d\chi$, as a result of the perturbation
should not be included. The reasoning for this is simple. If we assume that
magnetic phenomena are generated through the dynamo action, then we
inference that rotational energy is the source for the energy converted 
to the pure magnetic term. Since our models do not account for rotation,
we discard the final term in the above equation. Thus, the change in heat
to be considered in the stellar luminosity equation is 
\begin{equation}
dQ = c_PdT - \frac{\delta}{\rho}dP + \frac{P\delta\nu}{\rho\alpha\chi}d\chi.
\label{eq:heat}
\end{equation}
LS95 were quick to point out, that at the instant of any magnetic perturbation,
the change in heat due to magnetic effects should be exactly zero. 

Finally, from the observed change in heat comes the definition of the 
adiabatic gradient. Adiabaticity requires constant entropy, and therefore
no heat exchange, meaning
\begin{equation}
\nabla_{ad} = \thermt{\ln T}{\ln P}{S}{\chi} 
  = \frac{P\delta}{\rho T c_P}
\end{equation}
as it is in the non-magnetic case. Again, only in appearance; the actual
variables are altered by the introduction of a magnetic perturbation.


\subsubsection{Mixing-length Theory}
\label{sec:mlt}
Convection is determined to occur in regions where a given fluid parcel
is unstable to a small displacement in the radial direction. The primary
method of determining convective stability is to analyze the density of 
a generalized fluid parcel. Parcels that are less dense than their surroundings
will travel radially outward until they reach a height within the star at
which the surrounding fluid has the same density as the parcel. 

Upon reaching this point, the fluid parcel is assumed to fully mix with
its surroundings becoming indistinguishable from the rest of the fluid. 
Conversely, if a fluid element is more dense than surrounding fluid, it 
will sink down to a greater depth in the star, following the same trend
as a rising convective element. In either case, gravity is the restoring 
force. One assumption is that the fluid parcel is considered to 
always be in pressure equilibrium with its surroundings. 

The distance over which a fluid parcel travels before mixing is the well-known
``mixing-length.'' Mixing-length theory (MLT) has been well established
as a local means of prescribing convection for a one-dimensional stellar evolution code
\citep{vitense53,bv58}. At locations where various differences in prescriptions
of MLT occur, we will specify our assumptions.

Stability of a fluid parcel is determined by comparing the density of
the element to that of the surroundings
\begin{equation}
D\rho = \left[\left(\frac{d\rho}{dr}\right)_e - 
  \left(\frac{d\rho}{dr}\right)_s\right]\Delta r
\end{equation}
where $e$ and $s$ denote quantities of the fluid element under consideration
and the surroundings, respectively. To ensure stability, we require $D\rho > 0$, 
meaning, the element is stable to small radial displacements,
\begin{equation}
\left(\frac{d\rho}{dr}\right)_e - \left(\frac{d\rho}{dr}\right)_s > 0.
\label{eq:ddiff}
\end{equation}
As the fluid parcel is displaced radially, LS95 reminds us that we must 
consider how $\chi$ of the parcel reacts. If the initial $\chi$ of the 
parcel does not change as the element is displaced,
\begin{equation*}
\left(\frac{d\ln \chi}{dr}\right)_e = 0.
\end{equation*}
Conversely, if $\chi$ of the element is always equal to that of the 
surrounding material, there must be a flux of $\chi$ as the element is
displaced,
\begin{equation*}
\left(\frac{d\ln \chi}{dr}\right)_e = \left(\frac{d\ln \chi}{dr}\right)_s.
\end{equation*}
It is therefore advantageous to relate the spatial gradient of magnetic 
energy density of the parcel to that of the surroundings by introducing
a free parameter, $f$, such that
\begin{equation}
\left(\frac{d\ln \chi}{dr}\right)_e = f\left(\frac{d\ln \chi}{dr}\right)_s
\label{eq:magflux}
\end{equation}
where $f$ varies between 0 and 1. Later, we will attempt to eliminate this
free variable and set it to a physically realistic value.

Expanding the density differentials introduced in Equation~(\ref{eq:ddiff})
and multiplying through by the pressure scale height\footnote{Here, we 
reveal that we are basing our MLT formulation on the pressure scale height,
$H_{P} = -(dr/d\ln P)$, as opposed to the temperature scale height since 
we assume pressure equilibrium between the fluid parcel and its surroundings.} 
casts the stability criterion according to known differentials,
\begin{equation}
\delta \nabla_e - (1-f)\nu\nabla_{\chi} - \delta\tgrad > 0.
\label{eq:stab1}
\end{equation}
where we made use of a series gradient definitions, 
\begin{eqnarray}
\tgrad & \equiv & \left(\frac{d\ln T}{d\ln P}\right)_s \\
\nabla_e & \equiv & \left(\frac{d\ln T}{d\ln P}\right)_e \\
\nabla_{\chi} & \equiv & \left(\frac{d\ln\chi}{d\ln P}\right)_s = 
  \left(\frac{d\ln\chi}{dr}\right)_s \left(\frac{dr}{d\ln P}\right)_s.
\end{eqnarray}
Here, we have defined the temperature gradient of the surrounding plasma,
temperature gradient fluid element in question, and the magnetic energy
density gradient, respectively.

The magnetic energy density gradient turns out to affect the final form 
of the temperature gradient of the fluid parcel. In particular, the relation
between the gradient of the parcel and that of the surrounding fluid, given
by Equation~(\ref{eq:magflux}). If $f = 0$, then we find that complete 
adiabaticity holds, meaning $\nabla_{e}\rightarrow\nabla_{\rm ad}$. However,
for the case that $f\ne 0$, any heat transferred away from the parcel will
be in the form of magnetic energy ($dQ = d\chi$). Thus, from Equation~(\ref{eq:heat}),
\begin{equation}
0 = c_{P}\,dT_e - \frac{\delta}{\rho}dP_e + \frac{P\delta\nu}{\rho\alpha\chi}d\chi_e
\end{equation}
which may be rearranged to read
\begin{equation}
\nabla_e = \nabla_{\rm ad} - f\frac{\nu}{\alpha}\nabla_{\rm ad}\nabla_{\chi}. 
\end{equation}

Substituting this expression for the parcel's temperature gradient back
into Equation~(\ref{eq:stab1}), we derive the condition that must be met if a 
fluid parcel is to be stable against convection,
\begin{equation}
\nabla_{\rm ad} - f\frac{\nu}{\alpha}\nabla_{\rm ad}\nabla_{\chi} - 
     (1-f)\frac{\nu}{\delta}\nabla_{\chi} > \tgrad.
\label{eq:stab_crit}
\end{equation}
With a modified stability criterion in hand, LS95 the proceed to derive
a set of five equations that allow for a solution to the temperature 
gradient, $\tgrad$. The equations are developed through a detailed consideration
of the various energy fluxes through the system in their Sections 5.1 -- 5.3.
The final five equations comprising their magnetic mixing-length description
of convection are
\begin{eqnarray}
F_{\rm tot} & = & \frac{4acG}{3}\frac{T^4M_r}{\kappa P r^2}\nabla_{\rm rad} \label{eq:fluxrad}  \\
F_{\rm tot} & = & \frac{4acG}{3}\frac{T^4M_r}{\kappa P r^2}\tgrad + F_{\rm conv}  \\
F_{\rm conv} & = & \rho v_{\rm conv}\left[c_PDT + \frac{P\delta\nu}{\rho\alpha\chi}
               D\chi\right] \label{eq:conflux} \\
v_{\rm conv}^2 & = & \frac{g\ell_m^2\delta}{8H_P}\left[\left(\tgrad-\nabla_e\right) + 
              \frac{\nu}{\delta}(1-f)\nabla_{\chi}\right] \label{eq:vconv}
\end{eqnarray}
\begin{align}
 \frac{2acT^3}{\rho v_{\rm conv}c_P}\left[\frac{\omega}{1 + \ddot{y}\omega^2}\right]
           &(\tgrad-\nabla_e) \nonumber \\
  &  = (\nabla_e - \nabla_{\rm ad}) + f\frac{\nu}{\alpha}\nabla_{\rm ad}\nabla_{\chi} \label{eq:connect}
\end{align}
where we must now take a moment to dissect the various pieces. 

The first two
equations above describe the total flux if only radiation is carrying energy
and in the case that convection is also present, respectively. Within those 
two equations, $a$, $c$, and $G$ are, respectively, the radiation constant, 
speed of light in a vacuum, and the gravitational constant. $M_{r}$ is the 
mass contained within a spherical volume characterized by the radius, $r$,
and $\kappa$ is the radiative opacity. Lastly, $\nabla_{\textrm{rad}}$ is 
the radiative temperature gradient.

Equation~(\ref{eq:conflux}) is the energy flux transported by convection. 
The quantities, $DT$ and $D\chi$ are defined the same as $D\rho$ in 
Equation~(\ref{eq:ddiff}). Another variable, the convective velocity 
$v_{\textrm{conv}}$ is also introduced and characterizes the velocity of
the fluid within a convection cell.

The convective velocity is then defined in Equation~(\ref{eq:vconv}) and 
contains a single parameter not previously mentioned, $\ell_{m}$. This
parameter is the convective mixing-length, $\ell_{m}$, which is further
defined as some multiple of the pressure scale height (i.e., $\ell_m = 
\alpha_{\textrm{MLT}}H_{P}$). Note that the mixing-length introduces the canonical 
convective mixing-length parameter, $\alpha_{\textrm{MLT}}$, into the 
discussion.

Last of the five equations of MLT describes how the convective gradient 
is connected to the adiabatic gradient. Here, $\omega = \kappa\rho \ell_{m}$ and
$\ddot{y}$ is set by the geometry of the convecting bubble. The shape parameter,
$\ddot{y}$ is partially what separates the various formulations of MLT. 
Consistent with the standard DSEP treatment of convection, and LS95,
we set $\ddot{y} = 1/3$.

Combining Equations~(\ref{eq:fluxrad})--(\ref{eq:connect}) yields a solution
for the convective velocity, which effectively defines the temperature 
gradient. The entire solution, as with the previous derivations, may be found in its 
full glory in LS95. For our purposes, we cite yet another set of new 
variables required to simplify the solution,
\begin{equation}
Q = \delta
\end{equation}
\begin{equation}
\gamma_0 = \frac{c_P\rho}{2acT^3}\left(\frac{1 + \omega^2/3}{\omega}\right)
\end{equation}
\begin{equation}
C = \frac{g\ell^2_mQ}{8H_P}
\end{equation}
\begin{align}
 V^{-1} & = \gamma_0 C^{1/2}\cdot \nonumber \\
     & \left(\nabla_{\rm rad}-\nabla_{\rm ad} + 
        f\frac{\nu}{\alpha}\nabla_{\rm ad}\nabla_{\chi} + 
        (1-f)\frac{\nu}{\delta}\nabla_{\chi}\right)^{1/2}
\end{align}
\begin{equation}
A = \frac{9}{8}\frac{\omega^2}{\left(3 + \omega^2\right)}
\end{equation}
and, lastly,
\begin{equation}
y = v_{\rm conv}V\gamma_0.
\end{equation}
Resulting from the combination of the five magnetic MLT equations and the 
substitution of the variables just defined produces an equation that is 
quartic in $y$,
\begin{align}
\frac{2A}{V}y^4 + y^3 + \left[ 2A\gamma_0^2C \left(\frac{\nabla_{\rm ad}}{\alpha} - \frac{1}{Q}\right)(1-f)\nu\nabla_{\chi} + 1 \right] & Vy^2 - \nonumber \\
    y - \frac{C\gamma_0^2V^3\nu}{Q}(1-f)\nabla_{\chi} =  0. &
    \label{eq:longest}
\end{align} 
We remark that in the equation above, we deviate from LS95
by the presence of a $1/Q$ in the quadratic term. This difference appears
to be the result of the authors accidentally dropping a term in the original
derivation. However, as we shall see, this factor will become unimportant.

Finding a solution for $y$ from Equation~(\ref{eq:longest}) can easily be 
obtained numerically. Making an educated guess as to the solution for $y$
we may use a series of Newton--Raphson corrections to converge to a proper 
solution. A convergence tolerance of $10^{-10}$ is imposed on the correction term to 
reduce propagation of large numerical errors. Modifying the original ``guess''
supplied by LS95 to include the dropped term mentioned above, a good trial
solution is
\begin{equation}
y = \left[1 + 2AC\gamma_0^2\nu\left(\frac{\nabla_{ad}}{\alpha}-\frac{1}{Q}\right)(1-f)\nabla_{\chi}\right]^{-1}.
\end{equation}
One may notice that in the above trial solution (and most other MLT equations),
the precise value of the free parameter $f$ has the ability to drastically
simplify the expression.


\subsection{The Parameter $f$ and the Frozen Flux Condition}
\label{sec:ffc}

In Section~\ref{sec:mag_character} we specified that the plasma under consideration 
was perfectly conducting and, thus, had zero resistivity. One consequence 
of assuming an ideal MHD plasma is that magnetic field lines become 
physical objects that are transported by the plasma, the so-called frozen
flux condition \citep[FFC;][]{Alfven1942}. The magnetic flux is
\begin{equation}
\Phi(t) = \int_{S(t)} {\bf B}({\bf r}, t)\cdot d\hat{{\bf A}}.
\label{eq:bflux}
\end{equation}
For an ideal plasma, the evolution of $\Phi$ in time is dependent upon not
only the time rate of change of the magnetic field, ${\bf B}$({\bf $r$}, t), 
but also on any distortion occurring to the bounding surface, $S(t)$, as the
plasma moves. The net effect is that the time rate of change of the magnetic 
flux is equal to zero.  Therefore, we must have
\begin{equation}
\frac{\partial {\bf B}}{\partial t} = 0,
\end{equation}
which may be rewritten using Equation~(\ref{eq:ind}), the induction equation,
with $\eta = 0$. This results in the well-known FFC condition,
\begin{equation}
\nabla \times \left({\bf u} \times {\bf B}\right)= 0.
\end{equation}

The FFC enforces the restriction that, for a spherically 
symmetric bubble of plasma undergoing isotropic expansion or contraction
\citep{Kulsrud2004},
\begin{equation}
\frac{B}{\rho^{2/3}} = \textrm{constant}.
\end{equation}
Applying this constraint to the magnetic energy gradient definition of a 
convecting fluid element (Equation~(\ref{eq:magflux})), we are able to 
physically motivate the definition of the parameter $f$ which governs the 
flux of energy between the fluid element and the surrounding material.

Imagine a region in a star where a small bubble begins to grow convectively
unstable. Initially, the bubble has the same properties as the surrounding 
fluid. It is only because of the change in density that other properties
begin changing as well.
The FFC allows us to write the magnetic energy contained within 
a fluid parcel as a function of the magnetic energy of the surrounding
material. Since a convecting fluid parcel has a slight under- or overdensity
compared to its surroundings, we perturb the element's density 
\begin{equation}
\rho_{e} = \rho_{s} + \xi
\end{equation}
where it is understood that $\xi \ll \rho_{s}$. We also drop the subscript
$s$ hereafter. Assuming, for simplicity, that the bubble expands isotropically,
the magnetic field strength within a convectively unstable bubble is
\begin{equation}
B_{e} = \frac{B}{\rho^{2/3}} \left(\rho + \xi\right)^{2/3} = B\left(1 + \frac{\xi}{\rho}\right)^{2/3}
\end{equation}
meaning the magnetic energy per mass may be written as
\begin{equation}
\chi_{e} = \frac{B_e^2}{8\pi(\rho + \xi)} = \frac{B^2}{8\pi\rho}\left(1 + \frac{\xi}{\rho}\right)^{1/3}.
\end{equation}
We now have a direct relation between the magnetic energy density of the
convecting fluid element and the surrounding material,
\begin{equation}
\chi_{e} = \chi_s\left(1 + \frac{\xi}{\rho}\right)^{1/3}.
\end{equation}
Taking the radial, logarithmic derivative,
\begin{equation}
\left(\frac{d\ln \chi}{dr}\right)_e = \left(\frac{d\ln \chi}{dr}\right)_s + \frac{1}{3}\left(\frac{\xi}{\rho + \xi}\right)\left[\frac{d\ln\xi}{dr} - \frac{d\ln\rho}{dr}\right].
\end{equation}
The first of the bracketed terms goes to zero, as the density perturbation
is independent of radial location. Using the definition of $\chi_s$ (Equation~(\ref{eq:chi}))
to expand the derivative, and after rearranging the resulting terms, the 
equation becomes
\begin{equation}
\left(\frac{d\ln \chi}{dr}\right)_e = \frac{d}{dr}\ln\left(\frac{B^2}{8\pi}\right) - 
  \left(\frac{d \ln\rho}{dr}\right)
  \left[ 1 + \frac{1}{3}\left(\frac{\xi}{\rho + \xi}\right) \right].
\end{equation}
By definition, we know that $\xi/\rho \ll 1$, meaning the perturbation term
in the square brackets is negligible. As such,
\begin{equation}
\left(\frac{d\ln \chi}{dr}\right)_e \approx \left(\frac{d\ln \chi}{dr}\right)_s
\end{equation}
allowing us to conclude that the FFC implies that $f \approx 1$.


\subsection{Magnetic Field Strength Distribution}
\label{sec:mag_dist}
The strength of the perturbations described in the preceding sections
are determined by the magnitude and spatial gradient of $\chi$. Mentioned
in Section~\ref{sec:thermo}, was that we deviate from the prescription
of $\chi$ proposed by LS95. Instead of defining 
\begin{equation}
\chi = \chi_{\textrm{max}} \exp\left[ -\frac{1}{2}\left(\frac{M_D - M_{Dc}}{\sigma}\right)^2\right],
\end{equation}
where
\begin{equation}
M_D = \log_{10}\left[ 1 - \left(\frac{M_r}{M_{*}}\right) \right],
\end{equation}
we opt to directly prescribe the radial magnetic field profile. Approaching
the problem in this manner, however, introduces the difficulty of selecting 
a particular radial profile, and without any real confidence of the 
radial profile inside stars, we are left to our own devices. 

\begin{figure}
    \plotone{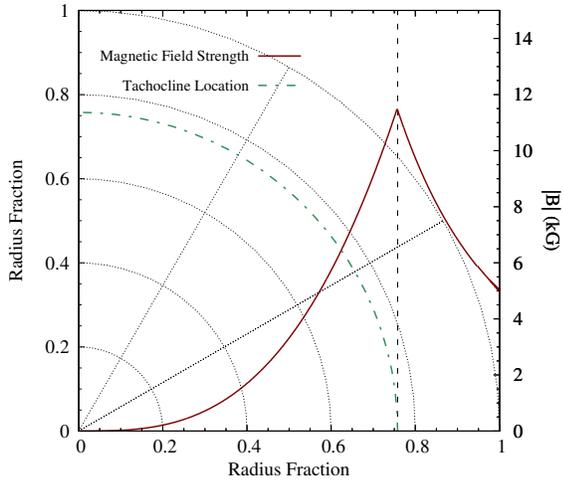}
    \caption{Magnetic field strength profile for a 1.0 $M_{\odot}$ star with
     a 5.0~kG photospheric magnetic field strength (maroon, solid).
     The green dash-dotted line indicates the location of the stellar tachocline,
     the interface between the radiative and convective regions. The plot is 
     meant only to illustrate the field strength profile. A small gap in the field 
     strength profile is barely perceptible near the surface of the star. This
     artifact is due to the separation of the stellar interior and envelope 
     integration regimes in the code.}
    \label{fig:mag_profile}
\end{figure}

One of the simplest profiles to select is that of a dipole configuration, where
the field strength drops off as $r^{-3}$ from the magnetic field source 
location. This is illustrated in Figure~\ref{fig:mag_profile}. The radial 
profile may then be prescribed as
\begin{equation}
B(R) = B(R_{\rm tach}) \cdot \left\{
    \begin{array}{l l}
        \left( R_{\rm tach}/R \right)^{3} & R > R_{\rm tach} \\
        \left( R/R_{\rm tach} \right)^{3} & R < R_{\rm tach}
    \end{array}
    \right.
\end{equation}
with the peak magnetic field strength defined to occur at the radius $R_{\rm tach}$.
The radial location described by $R_{\rm tach}$ is the location of 
the stellar tachocline, an interface between the convective envelope and
radiative core. This interface region is thought to be characterized
as a strong shear layer where the differentially rotating convection zone
meets the radiative core rotating as a solid body. Theory suggests that the 
tachocline is the source location for the standard mean-field stellar dynamo 
\citep[i.e., the $\alpha$--$\omega$ dynamo; ][]{Parker1975}.

Since DSEP monitors the shell location of the boundary to the convection
zone, the tachocline appeared to be a reasonable location, both theoretically
and numerically, to base the scaling of the magnetic field strength. However,
defining the magnetic field strength at the tachocline ($B(R_{\rm tach})$)
is required. In an effort to allow for direct comparisons between field 
strengths input into the code and observed magnetic field strengths, the
field strength at the tachocline is anchored to the photospheric (surface)
magnetic field strength,
\begin{equation}
B(R_{\rm tach}) = B_{\rm{surf}} \left( \frac{R_{*}}{R_{\rm tach}} \right)^{3}.
\end{equation}
where $B_{\textrm{surf}} = B(R_*)$ is introduced as a new free parameter. The advantage
of $B_{\rm surf}$ as a free parameter is that it has potential to be 
constrained observationally. 

Fully convective stars do not possess a tachocline. However, a dynamo 
mechanism still has the potential to drive strong magnetic fields through an
$\alpha^2$ mechanism \citep{CK06}. Full three-dimensional MHD modeling suggests that, in
this case, the magnetic field strength peaks at about 0.15~$R_{*}$
\citep{Browning2008}. Unfortunately, the adopted micro-physics were solar-like 
and may not be entirely
suitable for fully convective M-dwarfs. Regardless of these shortcomings,
Browning's investigation provides the only estimate, to date, for the location 
of the peak magnetic field strength in fully convective stars. As such, 
we adopt  $R_{\rm tach}=0.15 R_{*}$ as the dynamo source location in our models of fully
convective stars.


\subsection{Numerical Implementation}
Although we have laid out the mathematical construction of the magnetic 
perturbation, we have yet to illuminate precisely how the perturbation is
treated numerically. When a magnetic model is first executed, the user provides
a surface magnetic field strength, the geometry parameter $\gamma$, and the
age at which the magnetic perturbation will occur. The model proceeds to 
evolve the same as a standard model until the initial perturbation age is 
reached.

Once the perturbation age is reached, the magnetic field strength profile
is prescribed based on the assumed photospheric field strength and the location
of the tachocline, as in Figure~\ref{fig:mag_profile}. The magnetic energy 
and magnetic pressure are then
computed for each of the model's mass shells. Here, the total pressure associated
with each mass shell is also perturbed.

Following the introduction of the perturbation, the code must recompute 
the structure of the stellar envelope, which is separate from the stellar 
interior integration. The envelope comprises the outer 2\%--3\%, by 
radius, of the stellar model.
Surface boundary conditions are determined prior to the envelope
integration by interpolating within the {\sc phoenix} model atmosphere tables
using $\log g$ and $T_{\textrm{eff}}$. This interpolation returns $P_{\textrm{gas}}$
at the surface of the star, and defines the start of the inward integration.
The magnetic perturbation is then explicitly included in the calculation of the 
analytic EOS.

This leads into the convection routines, where the non-standard stability
criterion in Equation~(\ref{eq:stab_crit}) is evaluated and judged. Either 
the equations of magnetic MLT are solved, or the radiative gradient is selected.
The envelope integration scheme proceeds until it reaches a pressure commensurate 
with the pressure for the interior regime.

From the newly calculated envelope, the interior integration begins
using a Henyey integration scheme \citep{Henyey1964} with the magnetic 
perturbation implemented. The EOS and convection routines are evaluated as 
in the envelope. Once a final solution is converged upon, the code iterates
in time and the process is repeated.
For each temporal iteration, the magnetic field profile is adjusted to
adapt to changes in the location of the tachocline and changes in the number
of mass shells.


\begin{figure*}
    \plotone{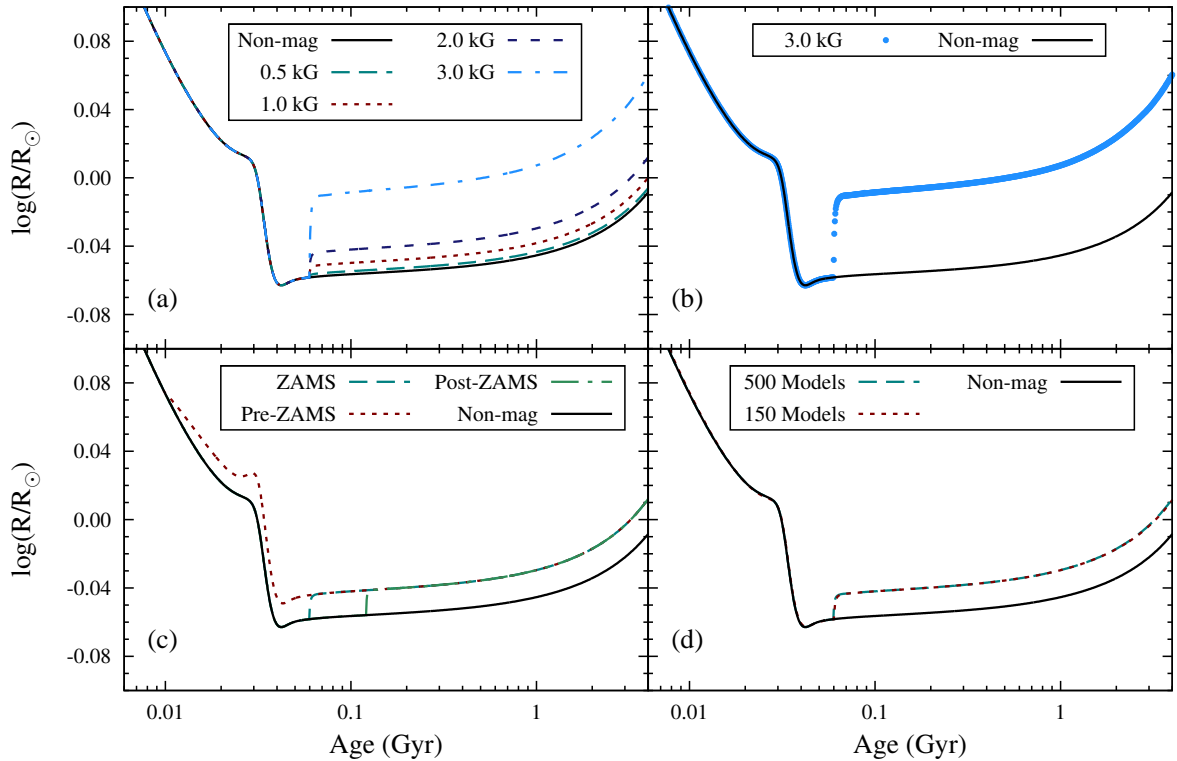}
    \caption{Tests of numerical stability for a 1 $M_{\sun}$ magnetic 
     model with $\gamma = 2$. (a) Influence of various magnetic
     field strengths. (b) Evidence of a smooth perturbation 
     at a large magnetic field strength with a relatively large radius change.
     (c) Consistency among models with the perturbation 
     turned on at various evolutionary stages. (d) 
     Demonstrating the insensitivity of the perturbation to the number of
     time steps after the perturbation is enabled.}
    \label{fig:num_test}
\end{figure*}


\section{Initial Testing}
\label{sec:num_test}
In Section~\ref{sec:mag}, we outlined the formulation and implementation
of a magnetic perturbation within the framework of DSEP. With the 
perturbation implemented, it was crucial to perform a series of numerical
tests and common-sense checks to validate that the code was operating properly.

The four key numerical tests were to ensure that:
\begin{enumerate}
    \item The relative change in radius between magnetic models of monotonically
     increasing photospheric magnetic field strength must also be monotonically 
     increasing with respect to a non-magnetic model.
    \item All model adjustments after the initial perturbation must be 
     continuous and smooth.
    \item The final perturbed model properties must be independent of the
     evolutionary stage at which the perturbation is made.
    \item The resulting perturbed model must be consistent, regardless of 
     the number of time steps taken after the perturbation.
\end{enumerate}
All of these tests were performed to confirm that the code was producing 
consistent results and that it was doing so in a numerically stable manner 
(i.e., no wild fluctuations).

Figure~\ref{fig:num_test} demonstrates that all model adjustments
to a magnetic perturbation satisfy each of the four criteria we required 
for numerical validation. Panel (a) 
demonstrates that the radius monotonically increases as the surface field
strength applied monotonically increases. Changes are observed to be smooth,
as seen in panel (b) and are independent of the number (or size)
of the evolutionary time steps taken (panel (d)). Finally, the plot in panel (c) indicates
that the relative change to the model asymptotes to the same value, regardless
of the evolutionary phase at which the perturbation is applied.

Beyond testing for numerical stability, we must be assured that the code 
produces results consistent with reality. Typically, a comparison with 
previous studies would be utilized. However, the only such examples computed
for evolutionary timescales are for CM~Dra \citep{Chabrier2007,MM12}. 
The stellar mass regime occupied by CM~Dra would require the implementation 
of FreeEOS, a task reserved for a future investigation. With the analytical
EOS, it would appear there are no models generated with which to compare.
Even LS95 operate over timescales on the order of a solar-cycle
and not evolutionary times. 

In the absence of previous results with which to compare, we opted to impose
a weak magnetic field (5~G) on our solar-calibrated model. Seeing as the 
properties of the Sun do not require a magnetic field in order to produce 
an adequate solar model \citep{Bahcall1997}, we would expect the solar properties
to remain almost entirely unaffected. As expected, the magnetically perturbed
model still meets our requirements for it to be considered solar-calibrated
(see Section~\ref{sec:std_model}). 
The model radius changes by 3 parts in $10^{5}$ while effective temperature 
changes by 4 parts in $10^{6}$ given the presence of the weak field.


\begin{figure*}
    \plottwo{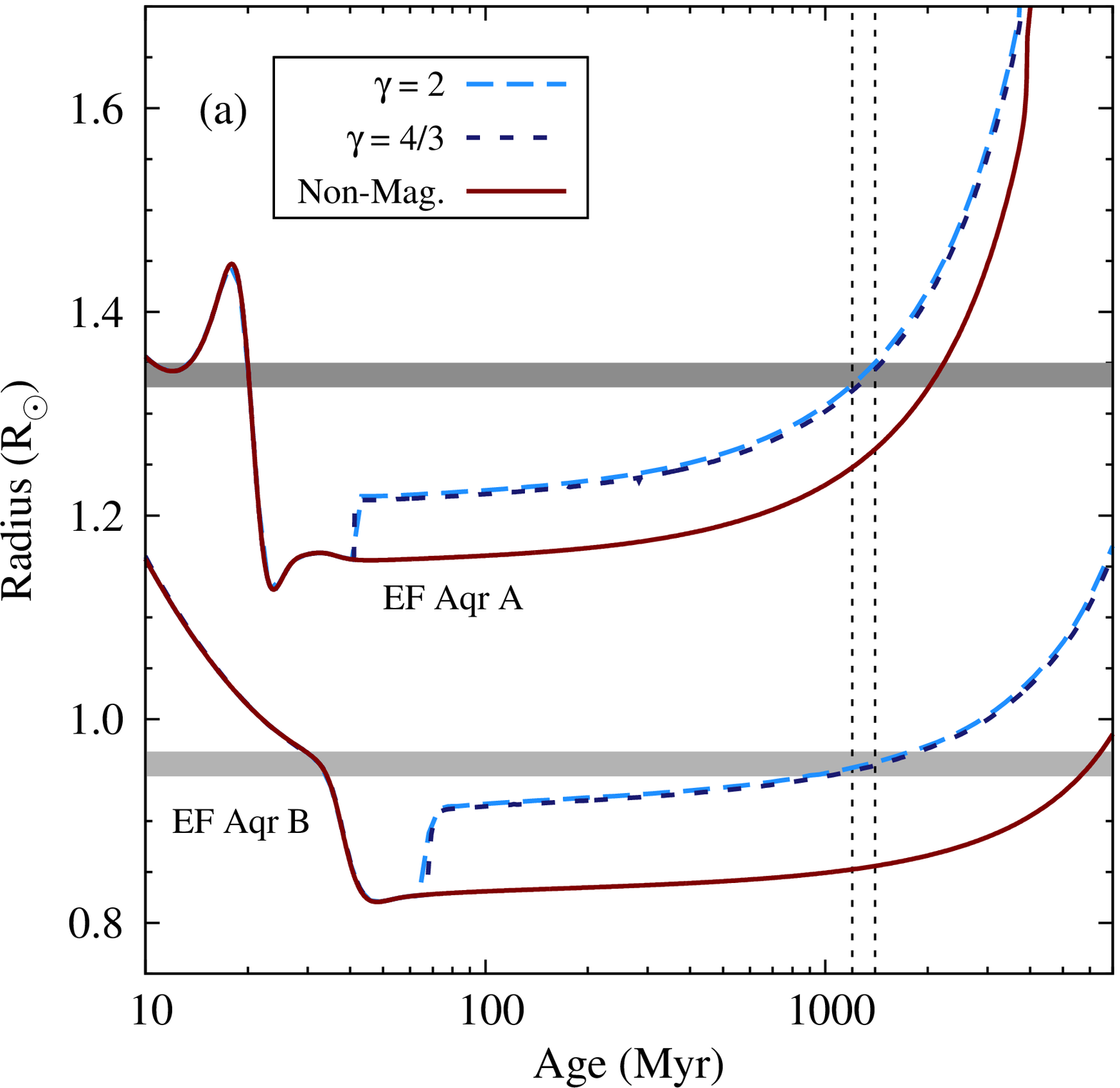}{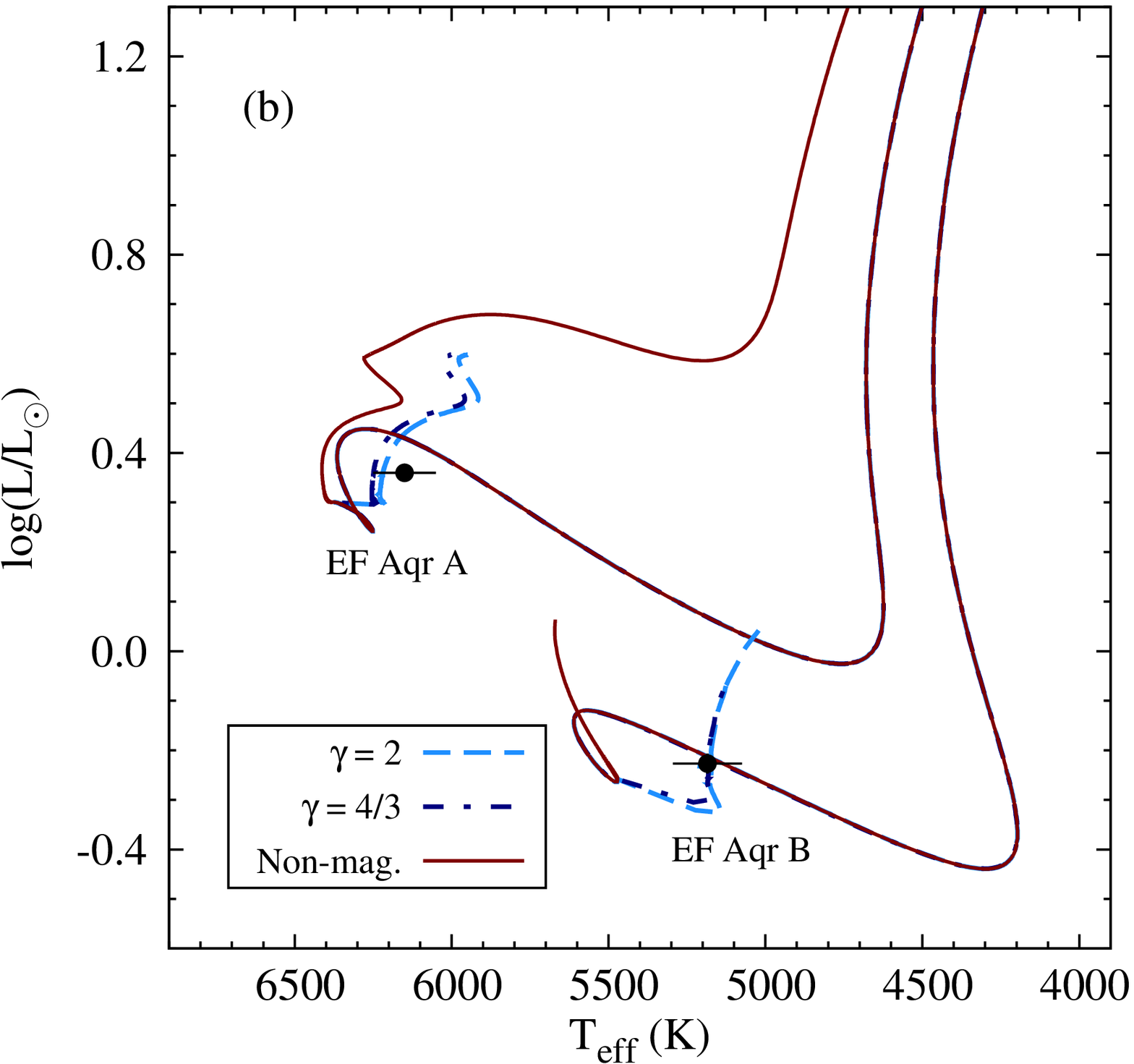}
    \caption{Individual stellar mass tracks representing EF~Aqr A and B (labeled). Non-magnetic mass tracks are shown as a red, solid line while $\gamma = 2$ and $\gamma = 4/3$ magnetic tracks are indicated by the blue, long-dash and light-blue, short-dash lines, respectively. The corresponding photospheric magnetic field strengths for EF Aqr A are 1.6~kG ($\gamma = 2$) and 2.6~kG ($\gamma = 4/3$). For EF Aqr B they are 3.2~kG ($\gamma = 2$) and 5.5~kG ($\gamma = 4/3$). (a) Age--radius diagram with the observed radii marked as gray regions. Vertical dotted lines define the age constraint imposed by EF Aqr A, suggesting the system has an age of $1.3\pm0.2$ Gyr. (b) HR diagram with the observed data marked as black points.}
    \label{fig:mag}
\end{figure*}


\section{Case Study: EF Aquarii}
\label{sec:results}

\subsection{Standard Models}
Standard, non-magnetic mass tracks with solar metallicity were computed for 
both EF~Aqr A and B with masses of 1.24~$M_{\odot}$ and 0.95~$M_{\odot}$, 
respectively. Additional mass tracks were also generated for a scaled-solar 
metallicity of +0.1 dex. The two components were unable to be fit with a coeval 
age, regardless of the adopted metallicity. This is consistent with the 
conclusions of \citet{vos12}. Two fitting methods were performed to this end. 

We first attempted to fit both components on an age--radius plane, which 
is equivalent to fitting on the mass--radius ($M$--$R$) plane. This is illustrated
by the solid lines in Figure~\ref{fig:mag}(a). The primary star
evolves much more rapidly than secondary, owing to the rather 
large mass difference. Thus, the model radius of the primary 
inflates to the observed radius at an age of 2.0~Gyr. The radius of the 
primary then quickly exceeds the observational bounds within about 
0.1~Gyr. The observational bounds on the primary radius places tight constraints
on the allowed age our models predict for the system. However, 
our model of the secondary does not reach the observed radius until an age of 
6.3~Gyr, an age difference of 4.3~Gyr between the two components. 
This 4.3 Gyr difference is consistently present in our models, including
when mass and composition are allowed to vary within the observed limits 
(not shown).

Assuming the age of the system was predicted accurately using models of EF~Aqr~A 
in the $M$--$R$ diagram, we find that the model radius of EF~Aqr~B underpredicts 
the observed radius by 11\%. Again, consistent with the findings of \citet{vos12}, 
who found a radius discrepancy of 9\%. Such a disagreement is also broadly 
consistent with results from other studies of active EBs \citep{Ribas2006,Torres2010}. 

A second approach was to fit the system on an HR diagram using the individual 
mass tracks. Figure~\ref{fig:mag}(b) demonstrates that the standard model tracks
do not match the observed $T_{\rm eff}$--luminosity of either star, despite 
fitting the stars individually in the $M$--$R$ plane. Our models predict temperatures 
that are 250~K (4\%) and 430~K (8\%) hotter than the observations for the 
primary and secondary, respectively.

The noticeable temperature disagreements in both stars may be the result 
of two possible effects. On the one hand, we might assume that 
the age predicted from the $M$--$R$ plane is correct and that there only exists 
a discrepancy with the effective temperatures. This implies that either the effective
temperature from the models or the observations is incorrect. However, since
the stars are quite similar to the Sun, it is more likely 
the case that our standard models underpredict the radius of the primary 
as well, driving up the model-derived effective temperature.
This particular scenario is supported by \citet{vos12} who found both 
components display obvious Ca {\sc ii} emission that is likely the result of 
each star having a magnetically heated chromosphere. 

The age inferred from the $M$--$R$ diagram
would then be older than the true age of the system. Unfortunately, this 
scenario further complicates the situation regarding EF~Aqr~B. If the age
of the system is younger than inferred from standard models of the primary,
then the radius discrepancy for the secondary becomes larger than 
originally quoted.

\subsection{Magnetic Models}
Following the results discussed above, magnetic models were computed for 
both EF~Aqr components using a scaled-solar heavy element composition. 
Several models with a mass of 1.24~$M_{\odot}$ were generated with various surface magnetic 
field strengths. We then found the model with the weakest field strength 
required to produce the observed radius and $T_{\textrm{eff}}$ of EF~Aqr~A. 
With $\gamma = 2$, we had to prescribe a photospheric magnetic field strength 
of 1.6~kG, while with $\gamma = 4/3$, a more intense 2.6~kG field was 
necessary. The magnetic model tracks are displayed in Figures~\ref{fig:mag}(a) 
and (b) as blue dashed lines. The magnetic models of the primary suggest 
a younger age of $1.3\pm0.2$~Gyr for the EF~Aqr system, as opposed to the 2.0~Gyr 
age determined from standard models. This younger age is consistent with 
the $1.5\pm0.2$~Gyr age derived for the primary by \citet{vos12} after 
fine-tuning the mixing length. 

We next had to select a magnetic field strength that would allow a 0.95~$M_{\odot}$ model 
to have a radius and $T_{\textrm{eff}}$ compatible with EF Aqr B at 1.3~Gyr, 
if finding that unique combination was possible. Surface magnetic 
field strengths of 3.2~kG and 5.5~kG were able to produce the required 
stellar parameters, with $\gamma$~=~2 and 4/3, respectively, at an age 
of 1.35~Gyr. In both cases, the models were able to reproduce the stellar 
radius and $T_{\textrm{eff}}$ within the quoted $1\sigma$ uncertainties. 
Figures~\ref{fig:mag}(a) and (b) demonstrate that the magnetic 
models do indeed match both component radii and $T_{\textrm{eff}}$s at a
common age. 

Structurally, the addition of a magnetic perturbation within the models 
reduces the radial extent of the surface convection zone. For both stars
in EF Aqr, we find the magnetic models that are sufficient to correct the
observed discrepancies have surface convection zones that are 4\% smaller
than those in the standard models, at the same age. The reduction of the 
surface convection zone can be attributed to the modified stability criterion
as well as modified convective velocities. While only speculation, we attribute 
the equality of the percent reduction of convection zone sizes to coincidence.


\section{Discussion}
\label{sec:disc}

\subsection{Field Strengths}
\label{sec:field_strength}
The implementation of a magnetic perturbation within stellar evolution models
is quite capable of reconciling predicted model fundamental stellar properties 
with those determined observationally, at least for EF~Aqr. While it seems 
plausible that magnetic fields may suppress thermal convection inside 
solar-type stars, how are we to be sure that magnetic fields may be reasonably 
invoked for this particular system? Even if invoking magnetic fields is 
rational, are the field strengths required by the models realistic? 

Addressing the first question, we showed in Section~\ref{sec:intro} that 
naturally inefficient convection, as described by the \citet{Bonaca2012}
calibration, was unable to account for the small values of $\alpha_{\rm MLT}$ 
required to mitigate the observed model-observation disagreements. But, 
the inability of naturally inefficient convection to provide a solution 
does not positively identify magnetic fields as the root cause. However, 
there is additional evidence that invoking magnetic fields is reasonable.

High-resolution spectroscopy of the Ca {\sc ii} H and K lines for both 
stars in EF Aqr reveals incredibly strong emission cores superposed on 
the absorption troughs \citep{vos12}. A search of the \emph{ROSAT} Bright
Source Catalogue \citep{Voges1999} also shows that EF~Aqr is a strong X-ray emitter. 
Coupling these observations with high projected rotational velocities
extracted from line broadening measurements, suggests that there is the
potential for a strong dynamo mechanism to be operating. This evidence is 
only circumstantial, but does provide tantalizing clues. 

For the sake of argument, let us assume that the stars are significantly 
magnetically active. It would then be worthwhile to compare the strength of the
magnetic field for each EF~Aqr component to those values required by the 
models. Unfortunately, no direct magnetic field strength estimates of EF~Aqr 
are available, forcing us to base our analysis on indirect magnetic field
strength estimates.

A natural first step would be to compare the EF~Aqr components to other 
known solar-type stars, such as the Sun and $\alpha$ Cen A and B (see 
Table~\ref{tab:mag}).
The Sun's mean photospheric magnetic field strength is between 0.1 G and 1 G
\citep{Babcock1955,Babcock1959,Demidov2002} with local patches of very intense fields 
(i.e., sunspots) with strengths on the order of 2--3 kG \citep{Hale1908}. 
Similarly, the average longitudinal field strength of $\alpha$ Cen A was determined
to be less than 0.2~G, after a null detection of a Stokes $V$ polarization 
signature \citep{Kochukhov2011}.
 
The field strengths required by the models of the EF~Aqr components therefore
suggest that the stars are pervaded by magnetic fields that typically characterize
the intense regions of sunspots. This at first appears detrimental to the validity of 
the models. However, studies of the active and quiet Sun, particularly 
sunspot regions, has led to multiple scaling relations allowing for an 
indirect determination of stellar photospheric magnetic field strengths.

One of these scaling relations was observed to exist between the X-ray
luminosity of an active region and its total unsigned magnetic flux
\citep{Fisher1998, Pevtsov2003}. The two observables were found to exhibit 
a power-law relation,
\begin{equation}
L_{x} \propto \Phi^{p},
\end{equation}
where the power-law index, $p$, was determined to be 1.19$\pm$0.04 by \citet{Fisher1998}.
The index was later revised by \citet{Pevtsov2003} using a more diverse
data set, including both solar and extrasolar sources.\footnote{The total 
unsigned magnetic flux of the stellar sources was obtained using direct
observational techniques \citep{Saar1996}.} Their revised analysis
decreased the index to $p=1.15$. The magnetic flux is defined in the usual manner
(Equation~(\ref{eq:bflux})),
\begin{equation}
\Phi = \int_{S} {\bf B}\cdot d\hat{{\bf A}} = \int_{S} \left|B_{z}\right| dA.
\end{equation}
with $|B_{z}|$ represents the vertical magnetic field strength. 
Therefore, if we are able to determine the X-ray 
luminosity of the EF~Aqr components, it is possible to place a lower limit 
on the magnetic field strength at the surface of the two components. 

The system has a confirmed X-ray counter-part in the \emph{ROSAT} All-Sky 
Survey Bright Source Catalogue \citep{Voges1999}. The X-ray count rate was 
converted to an X-ray flux according to the formula derived by \citet{Schmitt1995},
\begin{equation}
F_x = \left( 5.30 \textrm{HR} + 8.31\right)\times 10^{-12} X_{\textrm{cr}},
\end{equation}
where HR is the X-ray hardness ratio,\footnote{There are typically two hardness
ratios listed in the \emph{ROSAT} catalog, HR1 and HR2. The \citet{Schmitt1995} 
formula requires the use of HR1.} $X_{\rm cr}$ is the X-ray count rate, 
and $F_x$ is the X-ray flux. Finally, the X-ray flux was converted to a 
luminosity using the 172~pc distance quoted by \citet[][]{vos12}. 

The count rate measured by \emph{ROSAT} was 0.0655$\pm$0.0154 counts s$^{-1}$ with
a hardness ratio of 0.32$\pm$0.22. This yields an X-ray flux of $6.55\times10^{-13}$
erg~cm$^{-2}$. Weighting the contribution of each star to the total X-ray 
flux is reliably performed in one of two ways: by assuming both stars 
contribute equally \citep[valid if for binaries if stars are similar in 
radius;][]{Fleming1989} or weighting proportional to $v_{rot}\sin i$ 
\citep{Pallavicini1981,Fleming1989}.
Given the similarity of the two stars in EF Aqr, the precise weighting does
not affect the results. If both stars contribute equally to the total X-ray 
flux, then at a distance of 172~pc, the X-ray luminosity 
of each component is $L_{x}$~=~$1.16\times10^{30}$ erg~s$^{-1}$. Alternatively, 
weighting the two stars based on their projected rotational velocity gives
$L_{x,\,A}$~=~$1.25\times10^{30}$ erg~s$^{-1}$ and 
$L_{x,\,B}$~=~$1.07\times10^{30}$ erg~s$^{-1}$. 

To provide a comparison, the X-ray luminosity of a typical solar active 
region is on the order of $10^{27}$ erg~s$^{-1}$ \citep{Fisher1998, 
Pevtsov2003}. The Sun, on average, has a total X-ray luminosity of $10^{27}$
erg s$^{-1}$ up to nearly $10^{28}$ erg s$^{-1}$, depending on where in its
activity cycle it is located \citep{Ayers2009}. Similarly, \citet{Ayers2009}
monitored the X-ray luminosity of $\alpha$ Cen and found the primary had
an X-ray luminosity around half that of the Sun (approximately $10^{27}$
erg s$^{-1}$) and the secondary had about twice the X-ray luminosity of the
Sun (about $10^{28}$ erg s$^{-1}$). Further estimates for the X-ray luminosity
of $\alpha$ Cen A and B comes from \emph{ROSAT}, which yields luminosities 
between $10^{27}$ erg s$^{-1}$ and $10^{28}$ erg s$^{-1}$ for each component,
consistent with Ayers' analysis. Table~\ref{tab:mag} provides a comparison 
of how these quantities translate to magnetic field strengths.

Comparisons with the Sun and $\alpha$ Cen show that each component in EF 
Aqr has an X-ray luminosity 2--3 orders of magnitude greater than
``typical'' G and early-K stars. Again, while not indicative of causation, the 
correlation between high levels of X-ray emission and magnetic activity 
strongly suggests that EF Aqr is incredibly active. Given this information, 
our initial assumption that the stars are active seems valid. Therefore,
the implementation of a magnetic perturbation in stellar models of this
system appears warranted.

The amount of vertical magnetic flux near the surface of each star may then
be found using the \citet{Pevtsov2003} scaling relation. This suggests 
$\Phi = 1.39\times10^{26}$~Mx associated with each component (given equal
flux contribution). Converting to a magnetic field strength involves dividing the unsigned
magnetic flux by the total area through which the field is penetrating. 
For our purposes, the area is the entire surface area of the star.  
Therefore, we find the vertical magnetic field strength for the primary 
and secondary of EF~Aqr to be 1.3~kG and 2.5~kG, regardless of the adopted
flux weighting, respectively.
Note, this is the vertical magnetic field strength and sets a lower limit 
to the total magnetic field strength. It should also be mentioned that
the X-ray luminosities calculated
for EF Aqr A and B are near the edge of the data sample utilized by Pevtsov
et al., although no extrapolation of the relation was required. 

Further estimates of the magnetic field strengths may be found by applying
a scaling relation using the Ca {\sc ii} K line core emission \citep{Schrijver1989}.
The scaling relation was developed by correlating Ca {\sc ii} K line core 
emission and the magnetic flux density of solar active regions and their 
surroundings. The relation between the two was found to be
\begin{equation}
\frac{I_{c} - 0.13}{I_{w}} = 0.008\langle Bf \rangle^{0.6}
\end{equation}
where $I_c$ is the intensity of the emission line core, $I_w$ is the intensity
of the line wing. While the relation was derived for small, local active 
regions, \citet{Schrijver1989} suggest that there is no reason to believe that
the relation would not hold for hemispherical averages of solar-type stars.

Using the spectra provided by \citet{vos12} for the Ca~{\sc ii} K core
emission lines from each EF~Aqr component, we were able to estimate the magnetic field 
strength of each component. Spectra for the primary indicate that the average
magnetic field strength, $\langle Bf \rangle$, is equal to 830~G.
Similarly, for EF~Aqr~B, $\langle Bf \rangle$~=~3.3~kG. The values quoted 
above are derived from rough approximations of the core and wing intensities. However, 
we do not foresee the values of $\langle Bf \rangle$ changing radically with 
more precise line intensity measurements. We do caution that the results
for EF Aqr B require an extrapolation of the Ca {\sc ii} relation and the 
data for EF Aqr A place it near the edge of the derived relation where only
a few data points exist. 

Based on the scaling relations for X-ray emission and Ca~{\sc ii}~K line
core emission, the magnetic field strength for the primary and secondary
is seen to be approximately 1~kG and 3~kG, respectively. The magnetic 
field strengths required by the models are therefore within a factor of
two of the predicted field strengths, regardless of the adopted $\gamma$ 
value.  Since we expect the X-ray emission prediction to be a lower limit
to the full magnetic field strength, this is extremely encouraging. The models do 
not require abnormally large field strengths to reconcile the model properties
with those from observations, particularly when a $\gamma$ value of 2 is adopted.

\begin{deluxetable}{l c c c c}
    \tablewidth{\columnwidth}
    \tablecolumns{5}
    \tablecaption{Estimated Magnetic Field Strengths (in G)}
    \tablehead{
        \colhead{Star} & \colhead{Direct} & \colhead{X-Ray} & \colhead{Ca {\sc ii}}
        & \colhead{DSEP}
        }
    \startdata
    Sun             &  0.1 - 1.0  &  5 - 20   &  \nodata  &  \nodata      \\
    $\alpha$ Cen A  &  $< 0.2$    &  $\sim$3  &  \nodata  &  \nodata      \\
    $\alpha$ Cen B  &  \nodata   &  $\sim$49 &  \nodata  &  \nodata      \\
    EF Aqr A        &  \nodata   &  1300     &    830     &  1600 - 2600   \\
    EF Aqr B        &  \nodata   &  2500     &   3300     &  3200 - 5500
    \enddata
    \label{tab:mag}
\end{deluxetable}

\subsection{Implications}
The introduction of self-consistent magnetic stellar evolution models 
has multiple applications, ranging from studies of exoplanet host
stars \citep{Torres2007, gj1214, Muirhead2012} to investigations of 
cataclysmic variable (CV) donor stars \citep{Knigge2011}, as well as to
studies attempting to probe the stellar initial mass function of 
young clusters, where stars are typically very magnetically active 
\citep{JK2007,Jackson2009,Yang2011}. Although,
most obvious, are the implications for studies of low-mass eclipsing binary systems
\citep[see, e.g.,][and references therein]{Torres2010, Parsons2012}. 
Low-mass stellar evolution models have been highly criticized for being 
unable to predict the radii and effective temperatures of DEB stars. Models 
incorporating magnetic effects open the door to probing the underlying cause 
of the model-observation disagreements and providing semi-empirical 
corrections to models. 

Magnetic fields have long been theorized as the culprit, but previous 
generations of models have only treated magnetic fields in an ad hoc 
manner. Comparing the results of these methods with the one presented in 
this work, both in terms of surface parameters and the underlying interior
structure, will provide an 
interesting test of their validity. Ultimately, the ad hoc models
disagree on the dominant physical mechanism underlying
the observed discrepancies. The availability of self-consistent 
magnetic models should help to settle the debate as to which mechanism 
(suppressed convection or starspots) is most at work.

For EF~Aqr, the models suggest that magnetic suppression of thermal convection 
is sufficient to reconcile the models with the observations. Since
stars with small convective envelopes, such as those discussed in this work, 
are more sensitive to adjustments of the convective properties, it is not 
wholly surprising that suppressing convection is sufficient to explain the observations.
Whether this mechanism will be adequate for stars near the fully convective
boundary has yet to be seen. Future work modeling the lowest-mass DEB systems
will clarify this ambiguity. Regardless, this may suggest why the largest
radius deviations are predominantly observed at higher masses \citep{FC12}, 
with the notable exception of CM Dra \citep{Terrien2012}.

The nature of our models allows for independent verification of the 
magnetic field strengths required as input. While the indirect estimates 
provided by X-ray emission and Ca {\sc ii} K emission are encouraging, 
confirmation of these results using high resolution Zeeman spectroscopy, 
spectropolarimetry, or Zeeman--Doppler imaging (ZDI) is preferred. Unfortunately, 
these observations are difficult for fast rotators, such as those that 
comprise most DEB systems. They are also difficult for distant systems, where the short integration
time required by ZDI inhibits the ability of acquiring measurements with 
sufficient signal-to-noise
\citep[see the reviews by][]{Donati2009,Reiners2012}. Once a magnetic
field is detected, there exists the question of whether the observed strength
is indicative of the total magnetic field strength. Field strengths derived
for stars with spectral-type K and M using Stokes $V$ observations 
appear to yield only around 10\% of the total magnetic field strength 
compared to observations in Stokes $I$
\citep{Reiners2009}. This is a consequence of the fact that regions of 
opposite polarity tend to cancel out in Stokes $V$, making it most sensitive 
to the large-scale component, not the small-scale fields thought to pervade
low-mass stars. How
to accurately account for this when testing the models is not fully clear
and will require investigation. As more stars across all spectral types
are observed in both Stokes $V$ and $I$, a more coherent picture is sure to 
develop.

Magnetic models may also be useful for transiting exoplanet surveys,
particularly those focused on M-dwarfs \citep[e.g., MEarth transit survey;][]{NC08,
Irwin2009}. One of the largest uncertainties 
in deriving the properties of a transiting planet is the radius of the host star.
The lack of reliability involved in predicting low-mass stellar radii from
evolution models has deterred the use of models as predictors of exoplanet 
host-star radii \citep{Torres2007,gj1214}. Shoring up these deficiencies may 
lead to more accurate predictions of host star radii from stellar models,
circumventing, for a time, the need for lengthy and costly observations. This
would be most useful in identifying interesting follow-up targets by providing
a better estimate of the habitable zone \citep{Muirhead2012}.

There are certainly caveats with models, as other large uncertainties exist in predicting
the properties of a single star from stellar evolution mass tracks \citep{Basu2012}.
However, work is being performed to alleviate some of these uncertainties by
calibrating models to asteroseismic data \citep{Bonaca2012}. Low-mass 
stars are also less sensitive to the input parameters of stellar models
than their solar-type counterparts, reducing the associated uncertainties. 
Stellar evolution models may therefore
provide a fast and reliable estimate of the host star properties, depending
on the required level of precision.

Since most M-dwarfs being surveyed are nearby, there is a good chance that
they may have an X-ray counterpart in either the Bright Source or Faint Source 
Catalogue from the \emph{ROSAT} All-Sky Survey \citep{Voges1999,Voges2000}.
As was demonstrated in Section~\ref{sec:field_strength}, magnetic field strengths
required by the models are within about a factor of two (or better, if $\gamma
= 2$ is adopted) of those predicted
by the X-ray scaling relation of \citet{Pevtsov2003}. This will allow an
intelligent choice of the magnetic field strength used as an input for the
models, thus producing more reliable results from stellar models.

All told, the introduction of a self-consistent set of magnetic stellar 
evolution models provides the potential for models to be used with greater reliability
in a wide range of applications. There still exist several challenges that
require attention \citep{Boyajian2012}, but this is a first step in addressing
key issues that have been raised in the past two decades.


\acknowledgements
The authors express gratitude to L.~Hebb and O.~Kochukhov for helpful discussions.
G.A.F. thanks the William H. Neukom 1964 Institute for Computational 
Science for their generous support and the Department of Physics and Astronomy
at Uppsala University for their gracious hospitality. G.A.F. and B.C. also 
acknowledge the support of the National Science Foundation (NSF) grant 
AST-0908345. This research has made use of NASA's Astrophysics Data System, 
the SIMBAD database, operated at CDS, Strasbourg, France, and the \emph{ROSAT} 
data archive tools hosted by the High Energy Astrophysics Science Archive 
Research Center (HEASARC) at NASA's Goddard Space Flight Center.



\begin{thebibliography}{}
\bibitem[Alfv{\'e}n(1942)]{Alfven1942} Alfv{\'e}n, H.\ 1942, \nat, 150, 405
\bibitem[Ayers(2009)]{Ayers2009} Ayers, T.~R. \ 2009, \apj, 696, 1931
\bibitem[Babcock(1959)]{Babcock1959} Babcock, H.~W. \ 1959, \apj, 130, 364
\bibitem[Babcock \& Babcock(1955)]{Babcock1955} Babcock, H.~W. \& Babcock, H.~D. \ 1955, \apj, 121, 349
\bibitem[Bahcall et al.(2005)]{Bahcall2005} Bahcall, J.~N., Basu, S., Pinsonneault, M., \& Serenelli, A.~M. 2005, \apj, 618, 1049
\bibitem[Bahcall et al.(1997)]{Bahcall1997} Bahcall, J.~N., Pinsonneault, M., Basu, S., \& Christensen-Dalsgaard, J. \ 1997, Physical Review Letters, 78, 171
\bibitem[Baraffe et al.(1998)]{BCAH98} Baraffe, I., Chabrier, G., Allard, F., \& Hauschildt, P.~H. \
1998, \aap, 337, 403
\bibitem[Barnes(2010)]{Barnes2010} Barnes, S.~A. 2010, \apj, 722, 222
\bibitem[Basu et al.(2012)]{Basu2012} Basu, S., Verner, G.~A., Chaplin, W.~J., \& Elsworth, Y. \ 2012, \apj, 746, 76
\bibitem[Bender et al.(2012)]{Bender2012} Bender, C.~F., Mahadevan, S., Deshpande, R., et al. \ 2012, \apjl, 751, L31
\bibitem[Berger et al.(2006)]{Berger2006} Berger, D.~H., Gies, D.~R., McAlister, H.~A., et al. 2006, \apj, 644, 475
\bibitem[Bjork \& Chaboyer(2006)]{Bjork2006} Bjork, S.~R. \& Chaboyer, B. 2006, \apj, 641, 1102
\bibitem[B\"{o}hm-Vitense(1958)]{bv58} B\"{o}hm-Vitense, E. 1958, Z. Astrophys., 46, 108
\bibitem[Bonaca et al.(2012)]{Bonaca2012} Bonaca, A., Tanner, J.~D., Basu, S., et al. 2012, \apjl, 755, L12
\bibitem[Boyajian et al.(2012)]{Boyajian2012} Boyajian, T.~S., von Braun, K., van Belle, G., et al. \ 2012, \apj, 757, 112
\bibitem[Browning(2008)]{Browning2008} Browning, M. 2008, \apj, 676, 1262
\bibitem[Carter et al.(2011)]{Carter2011} Carter, J.~A., Fabrycky, D.~C., Ragozzine, D., et al. 2011, Science, 331, 562
\bibitem[Chaboyer et al.(2001)]{Chaboyer2001} Chaboyer, B., Fenton, W.~H., Nelan, J.~E., Patnaude, D.~J., \& Simon F.~E. 2001, \apj, 562, 521
\bibitem[Chaboyer \& Kim(1995)]{Chaboyer1995} Chaboyer, B. \& Kim, Y.-C. 1995, \apj, 454, 767
\bibitem[Chabrier \& Baraffe (1997)]{CB1997} Chabrier, G. \& Baraffe, I. \ 1997, \aap, 327, 1039
\bibitem[Chabrier et al.(2007)]{Chabrier2007} Chabrier, G., Gallardo, J., \& Baraffe, I. 2007, \aap, 472, L17
\bibitem[Chabrier \& K\"{u}ker(2006)]{CK06} Chabrier, G. \& K\"{u}ker, M. 2006, \aap, 446, 1027
\bibitem[Charbonneau et al.(2009)]{gj1214} Charbonneau, D., Berta, Z.~K., Irwin, J., et al.\ 2009, \nat, 462, 891 
\bibitem[Demarque et al.(2004)]{Demarque2004} Demarque, P., Woo, J-H., Kim, Y.-C., \& Yi, S.~K. 2004, \apjs, 155, 667
\bibitem[Demidov et al.(2002)]{Demidov2002} Demidov, M.~L., Zhigalov, V.~V., Peshcherov, V.~S., \& Grigoryev, V.~M.\ 2002, \solphys, 209, 217 
\bibitem[Donati \& Landstreet(2009)]{Donati2009} Donati, J.~F. \& Landstreet, J.~D. \ 2009, \araa, 47, 333
\bibitem[Dotter et al.(2007)]{Dotter2007} Dotter, A., Chaboyer, B., Jevremovi\'c, D., et al. 2007, \aj, 134, 376
\bibitem[Dotter et al.(2008)]{Dotter2008} Dotter, A., Chaboyer, B., Jevremovi\'c, D., et al. 2008, \apjs, 178, 89
\bibitem[Doyle et al.(2011)]{Doyle2011} Doyle, L.~R., Carter, J.~A., Fabrycky, D.~C., et al. 2011, Science, 333, 1602
\bibitem[Feiden \& Chaboyer(2012)]{FC12} Feiden, G.~A. \& Chaboyer, B. \ 2012, \apj, 757, 42
\bibitem[Feiden et al.(2011)]{Feiden2011} Feiden, G.~A., Chaboyer, B., \& Dotter, A. 2011, \apjl, 740, L25
\bibitem[Ferguson et al.(2005)]{Ferguson2005} Ferguson, J.~W., Alexander, D.~R., Allard, F., et al. 2005. \apj, 623, 585
\bibitem[Fisher et al.(1998)]{Fisher1998} Fisher, G.~H., Longcope, D.~W., Metcalf, T.~R., \& Pevtsov, A.~A. 1998, \apj, 508, 885
\bibitem[Fleming et al.(1989)]{Fleming1989} Fleming, T.~A., Gioia, I.~M., \& Maccacaro, T. 1989, \apj, 340, 1011
\bibitem[Gough \& Tayler(1966)]{GT66} Gough, D.~O. \& Tayler, R.~J. 1966, \mnras, 133, 85
\bibitem[Grevesse \& Sauval(1998)]{GS98} Grevesse, N., \& Suaval, A.~J. 1998, \ssr, 85, 161
\bibitem[Guenther et al.(1992)]{Guenther1992} Guenther, D.~B., Demarque, P., Kim, Y.-C., \& Pinsonneault, M.~H. 1992, \apj, 387, 372
\bibitem[Gurnett \& Bhattacharjee(2005)]{GB2005} Gurnett, D.~A. \& Bhattacharjee, A. 2005, Introduction to Plasma Physics (Cambridge: Cambridge Univ. Press)
\bibitem[Hale(1908)]{Hale1908} Hale, G.~E. \ 1908, \apj, 28, 315
\bibitem[Hauschildt et al.(1999a)]{Hauschildt1999a} Hauschildt, P.~H., Allard, F., \& Baron, E. 1999a, \apj, 512, 377
\bibitem[Hauschildt et al.(1999b)]{Hauschildt1999b} Hauschildt, P.~H., Allard, F., Ferguson, J., Baron, E., \& Alexander, D.~R. 1999b, \apj, 525, 871
\bibitem[Henyey et al.(1964)]{Henyey1964} Henyey, L.~G., Forbes, J.~E., \& Gould, N.~L. 1964, \apj, 139, 306
\bibitem[Iglesias \& Rogers(1996)]{Iglesias1996} Iglesias, C.~A. \& Rogers, F.~J., 1996, \apj, 464, 943
\bibitem[Irwin et al.(2009)]{Irwin2009} Irwin, J., Charbonneau, D., Nutzman, P., \& Falco, E.\ 2009, in IAU Symp. 253, Transiting Planets, ed. F. Pont, D. Sasselov, \& M. Holman (Cambridge: Cambridge Univ. Press), 37 
\bibitem[Irwin et al.(2011)]{Irwin2011} Irwin, J.~M., Quinn, S.~N., Berta, Z.~K., et al. 2011, \apj, 742, 123
\bibitem[Jackson(1998)]{Jackson1999} Jackson, J.~D. 1998, Classical Electrodynamics (3rd ed.; Hoboken, NJ: Wiley)
\bibitem[Jackson et al.(2009)]{Jackson2009} Jackson, R.~J., Jeffries, R.~D., \& Maxted P.~F.~L. 2009, \mnras, 399, L89
\bibitem[Johns-Krull(2007)]{JK2007} Johns-Krull, C.~M. \ 2007, \apj, 664, 975
\bibitem[Knigge et al.(2011)]{Knigge2011} Knigge, C., Baraffe, I., \& Patterson, J.\ 2011, \apjs, 194, 28 
\bibitem[Kochukhov et al.(2011)]{Kochukhov2011} Kochukhov, O., Makaganiuk, V., Piskunov, N., et al.\ 2011, \apjl, 732, L19
\bibitem[Kraus et al.(2011)]{Kraus2011} Kraus, A.~L., Tucker, R.~A., Thompson, M.~I., Craine, E.~R., \& Hillenbrand, L.~A. \ 2011, \apj, 728, 48
\bibitem[Kulsrud(2004)]{Kulsrud2004} Kulsrud, R.~M. 2004, Plasma Physics for Astrophysics (Princeton, NJ: Princeton Univ. Press)
\bibitem[Lacy(1977)]{Lacy1977} Lacy, C.~H. 1977, \apj, 218, 444
\bibitem[L\'{o}pez-Morales(2007)]{lopezm2007} L\'{o}pez-Morales, M. 2007, \apj, 660, 732
\bibitem[Lydon \& Sofia(1995)]{ls95} Lydon, T.~J. \& Sofia, S. 1995, \apjs, 101, 357 
\bibitem[MacDonald \& Mullan(2012)]{MM12} MacDonald, J. \& Mullan, D.~J., 2012, \mnras, 421, 3084
\bibitem[Mathis \& Zahn(2005)]{MZ05} Mathis, S. \& Zahn, J.-P 2005, \aap, 440, 653
\bibitem[Morales et al.(2010)]{Morales2010} Morales, J.~C., Gallardo, J.~J., Ribas, I., et al. 2010, \apj, 718, 502
\bibitem[Morales et al.(2008)]{Morales2008} Morales, J.~C., Ribas, I., \& Jordi, C. 2008, \aap, 478, 507
\bibitem[Morales et al.(2009)]{Morales2009} Morales, J.~C., Ribas, I., Jordi, C., et al. \ 2009, \apj, 691, 1400
\bibitem[Muirhead et al.(2012)]{Muirhead2012} Muirhead, P.~S., Hamren, K., Schlawin, E., et al.\ 2012, \apjl, 750, L37
\bibitem[Mullan \& MacDonald(2001)]{MM01} Mullan, D.~J. \& MacDonald, J. 2001, \apj, 559, 353
\bibitem[Nutzman \& Charbonneau(2008)]{NC08} Nutzman, P., \& Charbonneau, D.\ 2008, \pasp, 120, 317 
\bibitem[Pallavicini et al.(1981)]{Pallavicini1981} Pallavicini, R., Golub, L., Rosner, R., et al. 1981, \apj, 248, 279
\bibitem[Parker(1975)]{Parker1975} Parker, E.~N. 1975, \apj, 198, 205
\bibitem[Parsons et al.(2012)]{Parsons2012} Parsons, S.~G., Marsh, T.~R., G{\"a}nsicke, B.~T., et al.\ 2012, \mnras, 420, 3281 
\bibitem[Pevtsov et al.(2003)]{Pevtsov2003} Pevtsov, A.~A., Fisher, G.~H., Acton, L.~W., et al. 2003, \apj, 598, 1387
\bibitem[Popper(1997)]{Popper1997} Popper, D.~M. \ 1997, \aj, 114, 1195
\bibitem[Reiners(2012)]{Reiners2012} Reiners, A.\ 2012, Living Rev. Solar Phys., 9, 1 
\bibitem[Reiners \& Basri(2009)]{Reiners2009} Reiners, A. \& Basri, G. \ 2009, \aap, 496, 787
\bibitem[Ribas(2006)]{Ribas2006} Ribas, I. 2006, \apss, 304, 89
\bibitem[Richard et al.(2005)]{Richard2005} Richard, O., Michaud, G., \& Richer, J. 2005, \apj, 619, 538
\bibitem[Saar(1996)]{Saar1996} Saar, S.~H.\ 1996, in IAU Colloq.~153: Magnetodynamic Phenomena in the Solar Atmosphere, Prototypes of Stellar Magnetic Activity, ed. Y. Uchida, T. Kosugi, \& H. S. Hudson (Dordrecht: Kluwer), 367
\bibitem[Schmitt et al.(1995)]{Schmitt1995} Schmitt, J.~H.~M.~M., Fleming, T.~A., \& Giampapa, M.~S.\ 1995, \apj, 450, 392 
\bibitem[Schrijver et al.(1989)]{Schrijver1989} Schrijver, C.~J., Cot\'{e}, J., Zwaan, C., \& Saar, S.~H.\ 1989, \apj, 337, 964 
\bibitem[Skumanich(1972)]{Skumanich1972} Skumanich, A. 1972, \apj, 171, 565
\bibitem[Skumanich et al.(1975)]{Skumanich1975} Skumanich, A., Smythe, C., \& Frazier, E.~N. \ 1975, \apj, 200, 747
\bibitem[Spada \& Demarque(2012)]{SD12} Spada, F. \& Demarque, P. \ 2012, \mnras, 422, 2255
\bibitem[Stassun et al.(2012)]{Stassun2012} Stassun, K.~G, Kratter, K.~M., Scholz, A., \& Dupuy, T.~J. \ 2012, \apj, 756, 47 
\bibitem[Terrien et al.(2012)]{Terrien2012} Terrien, R.~C., Fleming, S.~W., Mahadevan, S., et al. \ 2012, \apj, 760, L9
\bibitem[Thoul et al.(1994)]{Thoul1994} Thoul, A.~A., Bahcall, J.~N., \& Loeb, A. 1994, \apj, 421, 828
\bibitem[Torres(2007)]{Torres2007} Torres, G. 2007, \apjl, 671, L65
\bibitem[Torres et al.(2010)]{Torres2010} Torres, G., Andersen, J., \& Gim\'enez, A. 2010, A\&AR, 18, 67
\bibitem[Torres \& Ribas(2002)]{Torres2002} Torres, G. \& Ribas, I. \ 2002, \apj, 567, 1140
\bibitem[Vitense(1953)]{vitense53} Vitense, E. 1953, \zap, 32, 135
\bibitem[Voges et al.(1999)]{Voges1999} Voges, W., Aschenbach, B., Boller, Th., et al. \ 1999, \aap, 349, 389
\bibitem[Voges et al.(2000)]{Voges2000} Voges, W., Aschenbach, B., Boller, Th., et al. \ 2000, VizieR Online Data Catalog, 9029, 0 
\bibitem[Vos et al.(2012)]{vos12} Vos, J., Clausen, J.~V., J{\o}rgensen, U.~G., et al. 2012, \aap, 540, 64
\bibitem[Winn et al.(2011)]{Winn2011} Winn, J.~N., Albrecht, S., Johnson, J.~A., et al. \ 2011, \apjl, 741, L1
\bibitem[Yang \& Johns-Krull(2011)]{Yang2011} Yang, H. \& Johns-Krull, C.~M. \ 2011, \apj, 729, 83
\end{thebibliography}
\end{document}